\documentclass[a4paper,11pt]{article}
\pdfoutput=1 % if your are submitting a pdflatex (i.e. if you have
\usepackage{jheppub} % for details on the use of the package, please
\usepackage[T1]{fontenc} % if needed
\usepackage{amsthm,enumerate,float,ytableau}
\usepackage[utf8]{inputenc}
\usepackage{lmodern}
\newcommand{\al}{\alpha}
\newcommand{\be}{\beta}
\newcommand{\de}{\delta}

\newcommand{\vep}{\varepsilon}
\newcommand{\ga}{\gamma}
\newcommand{\ka}{\kappa}
\newcommand{\la}{\lambda}

\newcommand{\si}{\sigma}
\renewcommand{\th}{\theta}
\newcommand{\vp}{\varphi}

\newcommand{\ze}{\zeta}
\newcommand{\De}{\Delta}
\newcommand{\Ga}{\Gamma}

\newcommand{\Si}{\Sigma}

\let\wt\widetilde

\newcommand{\tc}{\tilde{c}}

\newcommand{\tell}{\widetilde{\ell}}
\newcommand{\tcL}{\widetilde{\cL}}

\newcommand{\tx}{\widetilde{x}}

\newcommand{\ssH}{\mathsf{H}}

\newcommand{\RR}{{\mathbb R}}

\newcommand{\cL}{{\mathcal L}}

\newcommand{\cN}{{\mathcal N}}

\newcommand{\noi}{\noindent}

\newcommand{\ket}[1]{|#1\rangle}
\newcommand{\bra}[1]{\langle#1|}

\newcommand{\mss}{\kern 1pt}

\renewcommand{\le}{\leqslant}
\renewcommand{\ge}{\geqslant}
\newcommand{\tends}[1]{\bbuildrel{\hbox to 2em{\rightarrowfill}}_{#1}^{}}

\newcommand{\csch}{\operatorname{csch}}

\newcommand{\arctanh}{\operatorname{arctanh}}

\newcommand{\sgn}{\operatorname{sgn}}

\newcommand{\tr}{\operatorname{tr}}

\newcommand{\diag}{\operatorname{diag}}

\newcommand{\iu}{\mathrm i}
\newcommand{\e}{\mathrm e}
\newcommand{\diff}{\mathrm{d}}

\newcommand{\sn}{\operatorname{sn}}

\newcommand{\en}{\enspace}

\newcommand{\Int}[1]{\,\mathop{\!#1}\limits^{\lower1ex\hbox{$\scriptstyle\circ$}}{}}

\newcommand{\dc}{c^\dagger}

\newcommand{\laeff}{\la_{\text{eff}}}
\newcommand{\vepeff}{\vep_{\text{eff}}}

\theoremstyle{remark}
\newtheorem{remark}{Remark}

\def\clap#1{\hbox to 0pt{\hss#1\hss}}

\def\mathclap{\mathpalette\mathclapinternal}

\def\mathclapinternal#1#2{%
  \clap{$\mathsurround=0pt#1{#2}$}}
\begin{document}

\title{Entanglement entropy of inhomogeneous XX spin chains with algebraic interactions}

\author{Federico Finkel}
\author{and Artemio González-López}

\affiliation{Universidad Complutense de Madrid, Departamento~de Física Teórica\\ Facultad de
  Ciencias Físicas, Plaza de las Ciencias 1, 28040 Madrid, SPAIN}
\emailAdd{ffinkel@ucm.es}
\emailAdd{artemio@ucm.es}
\abstract{We introduce a family of inhomogeneous XX spin chains whose squared couplings are a
  polynomial of degree at most four in the site index. We show how to obtain an asymptotic
  approximation for the Rényi entanglement entropy of all such chains in a constant magnetic field
  at half filling by exploiting their connection with the conformal field theory of a massless
  Dirac fermion in a suitably curved static background. We study the above approximation for three
  particular chains in the family, two of them related to well-known quasi-exactly solvable
  quantum models on the line and the third one to classical Krawtchouk polynomials, finding an
  excellent agreement with the exact value obtained numerically when the Rényi parameter~$\al$ is
  less than one. When~$\al\ge1$ we find parity oscillations, as expected from the homogeneous
  case, and show that they are very accurately reproduced by a modification of the
  Fagotti--Calabrese formula. We have also analyzed the asymptotic behavior of the Rényi
  entanglement entropy in the non-standard situation of arbitrary filling and/or inhomogeneous
  magnetic field. Our numerical results show that in this case a block of spins at each end of the
  chain becomes disentangled from the rest. Moreover, the asymptotic approximation for the case of
  half filling and constant magnetic field, when suitably rescaled to the region of non-vanishing
  entropy, provides a rough approximation to the entanglement entropy also in this general case.}
\maketitle
\flushbottom

\section{Introduction}\label{sec.intro}

The entanglement entropy of spin chains of XX type ---or, equivalently, systems of free spinless
fermions with nearest-neighbors hoppings--- has been intensively studied since the seminal work of
Jin and Korepin~\cite{JK04} for the homogeneous chain. Indeed, the bipartite entanglement entropy
of one-dimensional models is a convenient indicator of their criticality. The reason is that in
their critical phase these models are effectively described at low energies by a
($1+1$)-dimensional conformal field theory (CFT), whose entanglement entropy has been shown to
scale logarithmically with the block size~$L$~\cite{CC04JSTAT,CC09}. In fact, a fundamental
property of all XX-type spin chains is the fact that their entanglement entropy can be expressed
in terms of the eigenvalues of a (truncated) correlation matrix. In the homogeneous case this
matrix is Toeplitz (for closed chains) or Toeplitz$+$Hankel (for open ones), which makes it
possible to apply proved instances of the (generalized) Fisher--Hartwig
conjecture~\cite{FH68,Ba79,DIK11} to rigorously derive the leading asymptotic behavior of the
entanglement entropy. In this way it was shown that the Rényi entanglement entropy $S_\al$ of the
homogeneous XX spin chain is asymptotically proportional to $\log L$ in the open, closed and
(semi)infinite cases~\cite{JK04,CE10}, even for subsystems of more than one
block~\cite{CH09,CCT09,ATC10,FC10,CFGT17}. Moreover, the coefficient of $\log L$ in the asymptotic
formula for~$S_\al$ confirms that this model has central charge $c=1$, as expected.

In fact, corrections to the logarithmic behavior of $S_\al$ in the limit of large $L$ were
exhaustively analyzed by Calabrese and Essler for the closed homogeneous XX chain~\cite{CE10}, and
by Fagotti and Calabrese for the open one~\cite{FC11}. For $\al>1$, or $\al\ge 1$ in the open
case, these terms present an oscillatory behavior which is particularly simple in the open case.
Indeed, in this case the leading order correction is proportional to $\sin((2L+1)k_F)L^{-1/\al}$,
where $k_F$ is the Fermi momentum. Moreover, it is argued in Ref.~\cite{FC11} (and earlier
in~\cite{CCEN10}) that this correction ---more precisely, the exponent of $L$ in the formula for the
amplitude--- encodes additional information about the underlying CFT beyond the central charge.

The situation is less straightforward in the non-homogeneous case, since the correlation matrix is
in general neither Toeplitz nor Toeplitz$+$Hankel. However, at half filling and constant magnetic
field, the leading behavior of the bipartite entanglement entropy can be derived through the
technique first used in Ref.~\cite{RDRCS17} to study the rainbow chain~\cite{VRL10}. The main idea
is that with these assumptions the chain's continuum limit yields the CFT of a massless Dirac
fermion in a static curved ($1+1$)-dimensional spacetime, whose metric's conformal factor is
proportional to the square of (the continuum limit of) the hopping amplitude. This suggests that
in the thermodynamic limit the leading asymptotic behavior of $S_\al$ can be obtained from the
formula for the homogeneous case replacing the chain's and block's lengths by their conformal
versions~\cite{DSVC17}. This was actually shown to be the case for the rainbow chain in
Ref.~\cite{TRS18}, and more recently for several other inhomogeneous XX chains in
Refs.~\cite{FG20,MSR21}. We stress, however, that to the best of our knowledge the latter results
only hold in the case of half filling and constant magnetic field. Moreover, the method just
outlined has only been applied to the leading term in the asymptotic formula for $S_\al$, without
addressing the behavior of the subleading corrections.

Another fundamental property of XX spin chains is their close connection with classical orthogonal
polynomials. Indeed, the chain's single-particle Hamiltonian is represented in the position basis
by a real tridiagonal symmetric matrix (the so-called hopping matrix), whose elements can in turn
be used to define a three-term recursion relation determining a \emph{finite} orthogonal
polynomial system (OPS) $\{P_n\}_{n=0}^N$, where $N$ is the number of spins. This establishes a
one-to-one correspondence between XX spin chains and OPSs, that can be used to derive in a simple
way many of the chain's properties. Indeed, the single-particle energies are the roots of the
critical polynomial $P_N$, and the correlation matrix elements can be computed in closed form
(without need of numerical diagonalization) in terms of the polynomials in the OPS evaluated at
the latter energies. This turns out to be more efficient than brute force diagonalization of the
matrix of the single-particle Hamiltonian as the number of spins grows. This connection has also
been exploited in Ref.~\cite{CNV19} to construct in some cases a tridiagonal matrix commuting with
the hopping matrix of the entanglement Hamiltonian, which can be used to improve the numerical
accuracy of the eigenvalues of the latter matrix and hence of the entanglement entropy.

A key property shared by the chains studied in Ref.~\cite{CNV19} is the fact that the square of
the interaction strength $J_n$ is a polynomial of degree at most four in the site index $n$. This
property is very natural from the point of view of the associated orthogonal polynomial family,
since $-J_{n-1}^2$ coincides with the coefficient of $P_{n-1}$ in the recursion relation for
$P_{n+1}$. In fact, the latter property also holds for the inhomogeneous XX chains related to
one-dimensional quasi-exactly solvable (QES) models~\cite{Tu88,Sh89,ST89,Us94} introduced in
Ref.~\cite{FG20}. We shall show in this work that for \emph{all} XX spin chains for which $J_n^2$
is a polynomial of degree up to four in $n$ it is possible to evaluate in closed form the leading
asymptotic approximation to the ground-state entanglement entropy (at half filling and in a
constant magnetic field) by the procedure introduced in Ref.~\cite{RDRCS17}. This in done
essentially by reducing a suitable elliptic integral to Legendre canonical form, a procedure which
depends on the root pattern of $J_n^2$.

We shall analyze in some detail three inhomogeneous XX chains with interactions $J_n$ of the
algebraic form described in the previous paragraph (\emph{algebraic interactions}, for short). Two
of these chains arise from well-known QES potentials, namely the sextic oscillator and the Lamé
periodic potential, while the third one, introduced in Ref.~\cite{CNV19}, is associated to the
Krawtchouk discrete orthogonal polynomial family. We first of all check that the leading term in
the asymptotic approximation to the entanglement entropy obtained through the related massless
Dirac fermion CFT in a suitably curved background is in excellent agreement with the numerical
results in the standard scenario of constant magnetic field and half filling.
%
% Changed
%
Moreover, for the sextic and Krawtchouk chains at half filling and in a constant magnetic field we
have found strong numerical evidence that the subleading (constant) term in the asymptotic
expansion of the entanglement entropy coincides with its counterpart for the homogeneous XX chain
for $N$ large enough. To the best of our knowledge, this remarkable coincidence had not been
previously noticed in the literature.
We stress in this regard that the connection of the models under study with families of orthogonal
polynomials makes it possible to determine the eigenvalues and eigenvectors of the single-particle
Hamiltonian in a numerically efficient way when the number of spins is very large. Our numerical
calculations also indicate that when $\al\ge1$ the Rényi entanglement entropy $S_\al$ features
parity oscillations which become more marked as $\al$ increases, as in the homogeneous case.
Remarkably, these oscillations are reproduced with great precision by the heuristic formula
proposed by Fagotti and Calabrese for the homogeneous XX chain, replacing the lengths of the block
and the whole chain by their values computed with the metric of the ambient space of the
associated CFT. More precisely, this formula depends only on two free parameters, whose fitted
values are very close to the theoretical ones for the homogeneous XX chain. This underscores the
essential similarity of the homogeneous and inhomogeneous cases, and the universality of Fagotti
and Calabrese's formula for this class of models.

All of the above results have been obtained under the standard assumptions of constant magnetic
field and half filling, for which the connection with the massless Dirac fermion CFT has a
theoretical justification via the continuum limit. In this work we have also analyzed the
non-standard situations of arbitrary filling and/or inhomogeneous magnetic field, which to the
best of our knowledge have not been addressed in the literature. Our numerical calculations
clearly indicate that the new feature in both of these scenarios is the vanishing of the
entanglement entropy when the length of the block is small or close to the chain's length. In
other words, the first few and last spins become disentangled from the rest of the chain. As a
consequence, the (leading) asymptotic approximation to the entanglement entropy derived from the
associated CFT cannot be expected to hold in this case. Remarkably, we have checked that this
approximation roughly reproduces the average behavior of $S_\al$ if suitably scaled to the region
of non-vanishing entropy. On the other hand, the oscillations of $S_\al$ in these non-standard
cases are found to be much more complex than in the usual situation of half filling and constant
magnetic field, and in particular are not well reproduced by the conformally modified version of
the Fagotti--Calabrese formula.

The paper is organized as follows. In Section~\ref{sec.model} we briefly review the connection
between inhomogeneous XX spin chains and free fermion systems, and recall how the latter models
can be exactly diagonalized. Likewise, in Section~\ref{sec.OP} we explain how these models are
related to a finite orthogonal polynomial family through its three-term recursion relation, and
how to exploit this connection to diagonalize the single-particle Hamiltonian. In
Section~\ref{sec.ee} we summarize the main results on the bipartite entanglement entropy of spin
chains used throughout the paper. In particular, we review some known results about the asymptotic
behavior of the entanglement entropy of the homogeneous XX chain as the number of spins tends to
infinity, and briefly outline their recent extension to the non-homogeneous case at half filling
in a constant magnetic field. In Section~\ref{sec.alg} we present a class of inhomogeneous XX spin
chains with algebraic interactions for which it is possible to compute in closed form an
asymptotic approximation to the block entanglement entropy by the procedure explained above. The
next three sections are devoted to the detailed analysis of three models in the previous family,
associated to the QES sextic oscillator Hamiltonian (Section~\ref{sec.sextic}), the classical
Krawtchouk polynomials (Section~\ref{sec.Krawtchouk}) and the periodic Lamé potential
(Section~\ref{sec.Lame}). In Section~\ref{sec.conc} we present our conclusions and discuss several
lines for future research. The paper ends with a technical appendix explaining how to reduce to
Legendre canonical form the elliptic integral appearing in the asymptotic formula for the
entanglement entropy.

\section{Inhomogeneous XX spin chains and free fermion systems}\label{sec.model}

The Hamiltonian of an inhomogeneous XX spin chain with interactions $J_n$ in an external magnetic
field $B_n$ can be taken as
\begin{equation}
  \label{Hchain}
  H = \frac12\sum_{n=0}^{N-2}J_n(\si_n^x\si_{n+1}^x+\si_n^y\si_{n+1}^y)
  +\frac12\sum_{n=0}^{N-1}B_n(1-\si_n^z)\,,
\end{equation}
where $N$ is the number of sites and $\si_n^\al$ (with $\al=x,y,z$) denotes the Pauli matrix
$\si^\al$ acting on the $n$-th site. In what follows we shall assume that the interaction
strengths $J_n$ do not vanish. As remarked in Ref.~\cite{FG20}, the model with $J_n$ replaced by
$\vep_nJ_n$, where $\vep_n\in\{\pm1\}$ is a site-dependent sign, is unitarily equivalent to the
original one. Hence we can take all the $J_n$'s to be positive without loss of generality.

It is well known that the Jordan--Wigner transformation
\begin{equation}\label{WJ}
  c_n=\prod_{k=0}^{n-1}\si_k^z\cdot\si_n^{+}\,,\qquad 0\le n\le N-1\,,
\end{equation}
where $\si^{\pm}_n:=(\si^x_n\pm\iu\si^y_n)/2$, maps the Hamiltonian~\eqref{Hchain} into that of a
system of $N$ hopping spinless fermions,
\begin{equation}
  \label{Hf}
  H=\sum_{n=0}^{N-2}J_n(\dc_nc_{n+1}+\dc_{n+1}c_n)+\sum_{n=0}^{N-1}B_n\dc_n c_n\,.
\end{equation}
Here $\dc_n$ (resp.~$c_n$) is the operator creating (resp.~destroying) a fermion at site $n$, and
the coefficients $J_n$ and $B_n$ respectively represent the hopping amplitude and the chemical
potential of the fermions. In what follows we shall mainly deal with the free fermion
system~\eqref{Hf}, our results being easily translated to its spin chain
equivalent~\eqref{Hchain}.

In the homogeneous case (i.e., when $J_n$ and $B_n$ are site independent), the
Hamiltonian~\eqref{Hf} commutes with the translation operator along the chain sites and is thus
diagonal in momentum space. In the non-homogeneous case this symmetry is lost, but the Hamiltonian
can still be diagonalized by introducing suitable modes. More precisely, let
\begin{equation}\label{ssH}
  \ssH=(H_{nm})_{n,m=0}^{N-1}\,,\qquad\text{with}\quad H_{nm}=\bra n H\ket m=J_{n}\de_{m,n+1}+
  J_{n-1}\de_{m,n-1}+B_n\de_{nm}\,,
\end{equation}
denote the matrix of the restriction of $H$ to the single-particle sector with respect to the
position basis
\[
  \Big\{\ket n:=\dc_n\ket{\mathrm{vac}}\mid 0\le n\le N-1\Big\}\,,
\]
where $\ket{\mathrm{vac}}$ is the fermionic vacuum. Since the hopping matrix $\ssH$ is real and symmetric, it can be diagonalized
by a real orthogonal matrix $\Phi=(\Phi_{nk})_{n,k=0}^{N-1}$, namely
\begin{equation}\label{HPhi}
  \Phi^T\ssH\Phi=\diag(\vep_0,\dots,\vep_{N-1})\,,
\end{equation}
where $\vep_0<\dots<\vep_{N-1}\in\RR$ are the eigenvalues of~$\ssH$. Note that, since $\ssH$ is
tridiagonal with nonzero off-diagonal entries, all its eigenvalues are simple. Let us then define
a new set of fermionic operators through the relation
\begin{equation}\label{mommodes}
  \tc_k:=\sum_{n=0}^{N-1}\Phi_{nk}c_n\,,\qquad 0\le k\le N-1,
\end{equation}
which satisfy the canonical anticommutation relations (CAR) on account of the unitary (real
orthogonal) character of $\Phi$. It is easily shown that the Hamiltonian~\eqref{Hf} can be written
as
\begin{equation}\label{Hdiag}
  H=\sum_{k=0}^{N-1}\vep_k\tc^\dagger_k\tc_k\,,
\end{equation}
and is thus diagonal in the basis consisting of the states
\begin{equation}\label{mombasis}
  \tc^\dagger_{k_0}\cdots\tc_{k_l}^\dagger\ket{\mathrm{vac}}\,,\qquad 0\le k_0<\cdots<k_l\le N-1\,,
\end{equation}
whose corresponding energy is given by
\begin{equation}\label{spectrum}
  E(k_0,\dots,k_l)=\sum_{j=0}^{l}\vep_{k_j}\,.
\end{equation}
In particular, the one-particle eigenstates $\tc^\dagger_k\ket{\mathrm{vac}}$ (with $0\le k\le N-1$) represent
single-fermion excitation modes with energy $\vep_k$.

The case in which the magnetic field $B_n$ vanishes for all $n$ deserves special attention.
Indeed, in this case $H$ is equivalent to $-H$ under the unitary transformation
$c_n\mapsto(-1)^nc_n$ (which obviously preserves the CAR), so that the spectrum is symmetric about
zero:
\[
  \vep_{N-k-1}=-\vep_k\,,\qquad 0\le k\le N-1\,.
\]
This implies that the system possesses particle-hole symmetry, since if
\hbox{$n_1'<\cdots<n_{N-k-1}'$} with
$\{n_1',\dots,n_{N-k-1}'\}\cup\{n_1,\dots,n_k\}=\{0,\dots,N-1\}$ we have
\[
  E(n_1',\dots,n_{N-k-1}')=\sum_{i=0}^{N-1}\vep_i-E(n_1,\dots,n_k)=-E(n_1,\dots,n_k)\,.
\]
Moreover, from the equivalence of $H$ to $-H$ under $c_n\mapsto(-1)^nc_n$ we immediately obtain
the relation
\begin{equation}\label{Phisym}
  \Phi_{n,N-k-1}=(-1)^n\Phi_{nk}
\end{equation}
up to an $n$-independent sign. Thus when $N$ is even the ground state is the half-filled state
\begin{equation}\label{halff}
  \tc^\dagger_0\cdots\tc^\dagger_{N/2-1}\ket{\mathrm{vac}}\,,
\end{equation}
with Fermi momentum
$\pi/2$ and energy
\begin{equation}\label{E0N}
  E_0:=\sum_{k=0}^{\mathclap{N/2-1}}\vep_k\,.
\end{equation}
(When $N$ is odd the ground state is doubly degenerate, since the zero energy mode does not change
the total energy.)

\section{Orthogonal polynomials}\label{sec.OP}

As explained in the previous section, the diagonalization of the full Hamiltonian~\eqref{Hf} of a
free fermion system is achieved by diagonalizing the hopping matrix~$\ssH$ in Eq.~\eqref{ssH}.
Since the latter matrix is tridiagonal, it can be used to define a \emph{finite} orthogonal
polynomial system $\{\phi_n(E)\}_{n=0}^{N}$ through the three-term recursion relation
\begin{equation}\label{rrphi}
 E\phi_n(E)=J_n\phi_{n+1}(E)+B_n\phi_n(E)+J_{n-1}\phi_{n-1}(E)\,,\qquad
  0\le n\le N-1
\end{equation}
(with $\phi_{-1}:=0$). It is easily shown that the polynomial $\phi_N(E)$ is proportional to the
characteristic polynomial of $\ssH$, and that the matrix elements of $\Phi$ can be taken as
$\Phi_{nk}=\phi_n(\vep_k)$, where for each $k$ the constant $\phi_0(\vep_k)$ is determined up to a
sign by the normalization condition
\[
  \sum_{n=0}^{N-1}\phi_n^2(\vep_k)=1\,.
\]
Recall that the eigenvalues~$\vep_k$ of $\ssH$ are non-degenerate (i.e., the roots of $P_N$ are
simple), and hence the orthogonality relations
\[
\sum_{n=0}^{N-1}\phi_n(\vep_k)\phi_m(\vep_k)=0\,,\qquad 0\le m\ne n\le N-1\,,
\]
are automatically satisfied. As is customary, we shall work in what follows with the \emph{monic}
polynomial family $\{P_n(E)\}_{n=0}^{N}$, where $P_n$ is the unique monic polynomial proportional
to $\phi_n$. From Eq.~\eqref{rrphi} it follows that
\[
  P_n=\frac{\phi_n}{\phi_0}\,\prod_{k=0}^{n-1}J_k\,,
\]
and that the polynomials $P_n$ satisfy the normalized recursion relation
\begin{equation}\label{rrP}
  P_{n+1}(E)=(E-B_n)P_n-a_nP_{n-1}\,,\qquad 0\le n\le N-1\,,
\end{equation}
with $P_{-1}:=0$ and $a_n=J_{n-1}^2>0$\,. Conversely, a monic polynomial OPS defined by a
recursion relation of the form~\eqref{rrP} with $a_n>0$ determines a free fermion
system~\eqref{Hf} with hopping $J_n=\sqrt{a_{n+1}}$ and chemical potential $B_n$. From the
previous argument it follows that the one-particle energies $\vep_k$ are the roots of the critical
polynomial $P_N$. It is also shown in Ref.~\cite{FG20} that the entries $\Phi_{nk}$ of the real
orthogonal matrix $\Phi$ determining the mode creation/annihilation operators through
Eq.~\eqref{mommodes} can be taken as
\begin{equation}
  \label{phiP}
  \Phi_{nk}=\sqrt{\frac{w_k}{\ga_n}}\,P_n(\vep_k)\,,
  \qquad 0\le k,n\le N-1\,,
\end{equation}
where
\begin{equation}\label{gaw}
  \ga_n:=\prod_{k=1}^na_k\,,\qquad w_k:=\frac{\ga_{N-1}}{P_{N-1}(\vep_k)P_N'(\vep_k)}\,,\qquad 0\le
  k\le N-1\,.
\end{equation}
Note that the orthogonality of the (real) matrix $\Phi$ follows directly from the fact that the
family $\{P_n\}_{n=0}^{N-1}$ is orthogonal with respect to the discrete measure
$\sum_{k=0}^{N-1}w_k\de(E-\vep_k)$, with square norm $\|P_n\|^2=\ga_n$ (see Ref.~\cite{FG20} for
the details). In the particular case in which $B_n$ vanishes for all $n$ the recursion
relation~\eqref{rrP} implies that $P_n$ has the parity of $n$, i.e., $P_n(-E)=(-1)^nP_n(E)$. It
then follows from the definition of $w_k$ that $w_{N-k-1}=w_k$, and hence
\[
  \Phi_{n,N-k-1}=\sqrt{\frac{w_{N-k-1}}{\ga_n}}\,P_n(\vep_{N-k-1})=
  \sqrt{\frac{w_k}{\ga_n}}\,P_n(-\vep_{k})=(-1)^n\sqrt{\frac{w_k}{\ga_n}}\,P_n(\vep_k)=
  (-1)^n\Phi_{nk}\,,
\]
in agreement with Eq.~\eqref{Phisym}.

\section{Entanglement entropy}\label{sec.ee}

A quantitative measure of the entanglement entropy of a block~$A$ of spins of the
chain~\eqref{Hchain} ---or fermions in the system~\eqref{Hf}--- when the whole system is in a pure
state~$\ket\psi$ is the entropy of the block's density matrix
$\rho_A:=\tr_{\overline A}\ket\psi\bra\psi$, where the subindex in the trace operator indicates
that we are tracing over the degrees of freedom of the complementary set
$\overline A:=\{0,\dots,N-1\}\setminus A$. More precisely, we shall take $A=\{0,\dots,L-1\}$ and
work with the Rényi entropy $S_\al$ (where $\al>0$ is a real parameter) defined by
\[
  S_\al=\frac{1}{1-\al}\log\tr(\rho_A^\al)\,,
\]
whose $\al\to1$ limit $S_1$ is the usual von Neumann--Shannon entropy
\[
  S=-\tr(\rho_A\log\rho_A)\,.
\]
In general, the system's state~$\ket\psi$ shall be taken as an eigenstate~\eqref{mombasis} with
the first $M$ energy modes excited:
\begin{equation}\label{psiM}
  \ket{\psi_M}=\prod_{\mathclap{k=0}}^{\mathclap{M-1}}\tc^\dagger_{k}\,\ket{\mathrm{vac}}\,.
\end{equation}

An important property of the free fermion system~\eqref{Hf} is the fact that its eigenstates are
Gaussian. Thus the analogue of Wick's theorem can be applied to express the Rényi entanglement
entropy in terms of the eigenvalues $\nu_n$ ($n=0,\dots,L-1$) of the $L\times L$ correlation
matrix $C(L,M)\equiv C$ with entries
\[
  C_{nm}:=\bra{\psi_M}\dc_nc_m\ket{\psi_M}\,,\qquad n,m=0,\dots,L-1\,,
\]
through the formula
\[
  S_\al=\frac1{1-\al}\sum_{n=0}^{L-1}\log\Bigl[\nu_n^{\al}+(1-\nu_n)^\al\Bigr]
\]
or
\[
S=-\sum_{n=0}^{L-1}\Big[\nu_n\log\nu_n+(1-\nu_n)\log(1-\nu_n)\Big]
\]
for the von Neumann entropy (see, e.g., \cite{VLRK03,JK04,LR09}). Using Eq.~\eqref{mommodes} it is
straightforward to show that the correlation matrix can be computed from the matrix $\Phi$ ---or
equivalently, by Eq.~\eqref{phiP}, the OPS $\{P_n\}_{n=0}^{N}$--- as
\[
  C_{nm}=\sum_{k=0}^{M-1}\Phi_{nk}\Phi_{mk}\,,\qquad n,m=0,\dots,L-1\,.
\]
In other words, $C=\Phi_{LM}\Phi_{LM}^{\sf T}$, where $\Phi_{LM}$ is the matrix obtained by taking
the first $L$ rows and $M$ columns of $\Phi$.
\begin{remark}\label{rem.B}
  The state~$\ket{\psi_M}$ in Eq.~\eqref{psiM} can always be regarded as the ground state by
  adding to the Hamiltonian~\eqref{Hf} a homogeneous term~$-B\sum_{n=0}^{N-1}\dc_nc_n$ with
  $\vep_{M-1}<B<\vep_M$. This of course amounts to adding a multiple of the identity to the
  hopping matrix $\ssH$, but does not change its eigenvectors $(\Phi_{0k},\dots,\Phi_{N-1,k})$
  (with $k=0,\dots,N-1$), and thus leaves the matrix $\Phi$ and the correlation matrix $C$
  invariant. Since, as explained above, the entanglement entropy is determined by the eigenvalues
  of $C$, the entanglement entropy is also invariant.
\end{remark}

When $0<M/N<1$ the system~\eqref{Hf} is gapless, and is thus described at low energy by an
effective ($1+1$)-dimensional CFT. The asymptotic behavior of the entanglement entropy of such a
theory \emph{in Minkowski spacetime} was determined in Refs.~\cite{HLW94,CC04JSTAT}. More
precisely, when the spatial manifold is a finite interval~$[0,\cL]$ the entanglement entropy of a
subinterval~$[0,\ell]$ is given by
\begin{equation}\label{SCFT}
  S_\al=\frac{c}{12}\,(1+\al^{-1})\log\biggl(\frac{\cL}{\pi\De
    x}\sin\biggl(\frac{\pi\ell}{\cL}\biggr)\biggr)+c'_\al+o(1)\,,
\end{equation}
where $c$ is the central charge of the CFT, $c'_\al$ is a non-universal constant depending only on
the Rényi parameter $\al$, and $\De x$ is an ultraviolet cutoff. In particular, the leading
asymptotic behavior of $S_\al$ is entirely determined by the central charge~$c$, and is thus
universal.

For a homogeneous free fermion system~\eqref{Hf} the correlation matrix $C$ is ``Toeplitz plus
Hankel'', which makes it possible to derive the leading asymptotic behavior of $S_\al$ in the
limit $L,N\to\infty$ with~\cite{FC11}\footnote{An analogous result for the
  homogeneous XX chain with periodic boundary conditions, whose correlation matrix is simply
  Toeplitz, was derived earlier on by Jin and Korepin~\cite{JK04} using a proved case of the
  Fisher--Hartwig conjecture~\cite{Ba79}.}
\[
  \la:=\lim_{N\to\infty}\frac LN\in(0,1).
\]
At half filling this result is in agreement with the CFT formula~\eqref{SCFT} with central charge
$c=1$ taking $\De x$ as the chain's spacing, so that
\begin{equation}\label{cLell}
  \cL=(N-1)\De x\,,\qquad \ell=(L-1)\De x\,.
\end{equation}
This confirms the fact that in the homogeneous case the free fermion
system~\eqref{Hf} (which is equivalent to the homogeneous Heisenberg XX chain) is described by the
free fermion CFT (in Minkowski spacetime) with $c=1$.

In fact, the asymptotic behavior of the entanglement entropy of the homogeneous XX chain is known
in much greater detail~\cite{FC11}. To begin with,
in this case the non-universal constant
$c'_\al$ is given by
\begin{equation}\label{cpal}
  c'_\al=\frac12\,\big(1+\al^{-1}\big)\left\{\frac13\,\log2+\int_0^{\infty}\left[\frac{\csch
        t}{1-\al^{-2}}\left(\al^{-1}\csch(t/\al)-\csch t\right)-\frac{\e^{-2t}}6\right]\frac{\diff
      t}t\right\}
\end{equation}
for $\al\ne1$ (and its $\al\to1$ limit for $\al=1$~\cite{JK04}). Actually, Eq.~\eqref{SCFT} holds
for an arbitrary filling (with $c_\al'$ as above) if we add the extra factor $\sin k_F$ to the
argument of the logarithm, where $k_F:=\pi M/N$ is the Fermi momentum. For $\al<1$, the $o(1)$
term is actually of order $N^{-1}$, and thus the formula
\begin{equation}\label{SCFTkF}
  S_\al(N,\la)=\frac{c}{12}\,(1+\al^{-1})\log f_H(N,\la)+c'_\al+O(N^{-1})\,,
\end{equation}
with
\begin{equation}
  \label{fHdef}
  f_H(N,\la):=\frac{\cL}{\pi\De
    x}\sin\biggl(\frac{\pi\ell}{\cL}\biggr)\sin k_F\simeq \frac{N}\pi\,\sin(\pi\la)\sin k_F\,,
\end{equation}
provides an excellent approximation to the Rényi entanglement entropy of the homogeneous XX chain
even for moderately large values of $N$. On the other hand, for $\al\ge1$ the $o(1)$ term in
Eq.~\eqref{SCFT} features parity oscillations which for large $\al$ can even obscure the leading
asymptotic behavior~\eqref{SCFTkF}. A remarkable heuristic formula for these parity oscillations
was found in Ref.~\cite{FC11} using CFT arguments, namely
\begin{equation}\label{Sasymphom}
  S_\al(N,\la)=\frac1{12}\big(1+\al^{-1}\big)\log
  f_H(N,\la)+c_\al'+\mu_\al\sin\left((2L+1)k_F\right) f_H(N,\la)^{-1/\al}+o\bigl(N^{-1/\al}\bigr)\,,
\end{equation}
where $c'_\al$ is given by Eq.~\eqref{cpal} and
\begin{equation}\label{mutheor}
  \mu_\al=\frac{2^{1-\frac2\al}}{1-\al}\,
  \frac{\Ga(\frac12+\frac1{2\al})}{\Ga(\frac12-\frac1{2\al})}\,,\qquad\al\ne1,
\end{equation}
with $\mu_1=\lim_{\al\to1}\mu_\al=-1/4$. This formula reproduces with great precision the parity
oscillations of the Rényi entropy of the homogeneous XX chain when $\al\ge1$. It was also argued
in the latter reference that a subleading term proportional to $f_H(N,\la)^{-\ka/\al}$ ---though
not the coefficient $\mu_\al$ or even the oscillatory term $\sin\left((2L+1)k_F\right)$--- is in
fact \emph{universal}, the parameter $\ka$ (which is unity for the homogeneous XX chain) providing
information on the scaling dimensions of relevant operators in the associated CFT.

In the general (non-homogeneous) case the chain~\eqref{Hf} is no longer described by a CFT in
Minkowski spacetime, and thus the previous considerations do not directly apply. However, when $N$
is even and $B_n=0$ for all $n$ ---i.e., when the system's ground state is the half-filled
state~\eqref{halff}--- it was shown in Refs.~\cite{RDRCS17,TRS18} that the continuum limit of the
Hamiltonian~\eqref{Hf} coincides with the Hamiltonian of a free massless Dirac fermion in the
curved spacetime with static metric
\begin{equation}\label{Jmetric}
  \diff s^2=J(x)^2\diff t^2-\diff x^2\,.
\end{equation}
Here $J(x)$ is the continuum limit of $J_n$, obtained by setting $n\De x=:x_n$, taking the limit
$N\to\infty$ and $\De x\to0$ with $\cL=(N-1)\De x$ fixed, and replacing $x_n$ by a continuous
variable~$x\in[0,\cL]$. The metric~\eqref{Jmetric} can be expressed in isothermal coordinates as
\begin{equation}\label{isother}
  \diff s^2=J(x)^2(\diff t^2-\diff\tx^2),
\end{equation}
with $\diff\tx=\diff x/J(x)$. It is therefore natural to assume that in the limit $N\to\infty$
with $L/N\to\la$ finite the entanglement entropy of the free fermion system~\eqref{Hf} with even
$N$ and $B_n=0$ for all $n$ ---or more generally, by Remark~\ref{rem.B}, with $B_n$ constant at
half filling--- can be obtained from Eq.~\eqref{SCFT} with $\ell$, $\cL$ and $\De x$ respectively
replaced by the conformal lengths
\begin{equation}\label{tildels}
  \De\tx=\frac{\De x}{J(\ell)}\,,\qquad \tell=\tx(\ell)\,,\qquad\tcL=\tx(\cL)\,,
\end{equation}
where
\begin{equation}
  \label{tildex}
  \tx(x):=\int_0^x\frac{\diff s}{J(s)}
\end{equation}
is the length of the spatial interval~$[0,x]$ computed with the metric~\eqref{isother}. In other
words, we should have
\begin{equation}
  \label{Salasy}
  S_\al(N,\la)= \frac{1}{12}\,(1+\al^{-1})\log\left(\frac{\tcL}{\pi\De
    \tx}\sin\left(\frac{\pi\tell}{\tcL}\right)\right)+c_\al'+o(1)
\end{equation}
for a suitable (non-universal) constant $c_\al'$ (not necessarily given by Eq.~\eqref{cpal}). This
was shown to be the case for the rainbow chain (for which $J_n=J_0\e^{-h|n/N-1/2|}$,
$J(x)=J_0\e^{-h|x/\cL-1/2|}$) in Refs.~\cite{RDRCS17,TRS18}, and more recently for the
Lamé~\cite{FG20}, Rindler and sine chains~\cite{MSR21}.

An interesting open problem motivated by the previous considerations is whether the more precise
asymptotic approximations~\eqref{SCFTkF}--\eqref{Sasymphom} also hold in the non-homogeneous case
after the replacement $(\De x,\ell,\cL)\to(\De\wt x,\tell,\tcL)$ in Eq.~\eqref{fHdef}. In fact,
since the equivalence of the continuum limit of the inhomogeneous chain~\eqref{Hchain} with a CFT
in curved spacetime has only been established at half filling (and for zero magnetic field), the
latter formulas are only expected to apply when $k_F=\pi/2$ and $B_n$ vanishes (or, more
generally, is constant). We are thus led to conjecture the following more detailed asymptotic
formulas for the Rényi entanglement entropy of the general (non-homogeneous) chain~\eqref{Hchain}
at half filling in a constant magnetic field:
\begin{equation}\label{Sasymposc}
  S_\al=\frac1{12}\big(1+\al^{-1}\big)\log
  f(N,\la)+c_\al'+\begin{cases}O(N^{-1}),&\al<1\\
    \mu_\al (-1)^Lf(N,\la)^{-1/\al}+o\bigl(N^{-1/\al}\bigr),&\al\ge1\,,
  \end{cases}
\end{equation}
with
\begin{equation}
  \label{fnonh}
  f(N,\la):=\frac{\tcL}{\pi\De
    \wt x}\sin\biggl(\frac{\pi\tell}{\tcL}\biggr)\simeq
  \frac{N\tcL/\cL}{\pi}\,J(\ell)\sin\left(\frac{\pi\tell}{\tcL}\right)\,.
\end{equation}
In what follows we shall introduce a family of inhomogeneous XX chains for which it shall be
checked that the above conjecture holds. We shall also show that the analogous generalization of
Eqs.~\eqref{SCFTkF}--\eqref{Sasymphom} to the case of arbitrary filling and/or inhomogeneous
magnetic field is not valid for the models considered in this paper.

\section{Spin chains with algebraic interactions}\label{sec.alg}

In order to evaluate the right-hand side of Eqs.~\eqref{Sasymposc}-\eqref{fnonh} in closed form it
is necessary to compute the integral in Eq.~\eqref{tildex}. We shall introduce in this section a
large class of inhomogeneous XX chains for which the latter integral can be explicitly evaluated.
This class is characterized by the fact that the coefficient $a_n$ in the recursion
relation~\eqref{rrP} is a polynomial of degree at most four in $n$, and thus $J_n$ and $J(x)$ are
algebraic functions of degree two. As we shall discuss in the sequel, for these chains the RHS of
Eq.~\eqref{tildex} is an elliptic integral which can be evaluated by transforming it to Legendre
normal form. In fact, chains with this type of algebraic interactions have been recently discussed
in the literature in two different contexts. Indeed, the inhomogeneous chains associated to the
discrete Krawtchouk and dual Hahn polynomials studied in Ref.~\cite{CNV19}, whose entanglement
Hamiltonian admits a commuting tridiagonal operator, both feature interactions of the above form.
The same is true for all spin chains related to quasi-exactly solvable quantum models on the line
recently constructed and classified in Ref.~\cite{FG20}. In particular, it was shown in the latter
reference that the entanglement entropy of an inhomogeneous spin chain associated to the quantum
Lam\'e potential is indeed well approximated by the asymptotic formula~\eqref{Salasy} in the limit
of large $N$.

Consider, then, the integral~\eqref{tildex}. Since $J(x)$ is dimensionless (in natural units), it
must be a function of the dimensionless variable $\xi:=x/\cL$. Setting
\begin{equation}
  \label{pdef}
  J(x)=\sqrt{p(x/\cL)}
\end{equation}
we
can rewrite~\eqref{tildex} as
\begin{equation}\label{mainint}
  \tx(x)=\cL\int_0^{x/\cL}\frac{\diff\xi}{\sqrt{p(\xi)}}\,.
\end{equation}
We shall assume in what follows that $p$ is a polynomial with real coefficients, with
$\deg p\le 4$ and $p(\xi)\ge0$ for $0\le\xi\le1$. The main idea for reducing the last integral to
canonical form is the fact that a real projective change of variable
\begin{equation}\label{proj}
  \xi=\frac{a z+b}{c z+d}\,,\qquad\De:=ad-bc\ne0\,,
\end{equation}
transforms it into an integral of the same type. Indeed,
\[
  \int_0^{s}\frac{\diff\xi}{\sqrt{p(\xi)}}=\int_{z(0)}^{z(s)}\frac{\vep\,\diff z}{\sqrt{\hat
      p(z)}}\,,
\]
with $\vep:=\sgn\De$, $z(\xi)=(d\xi-b)/(a-c\xi)$ and
\[
  \hat p(z):=\frac{(cz+d)^4}{\De^2}\,p\Bigl(\tfrac{a z+b}{c z+d}\Bigr)\,
\]
a polynomial of degree at most four in $z$. Using a projective change of variable of the
form~\eqref{proj}, the original polynomial $p(\xi)$ can always be transformed into a suitable
canonical form~$\hat p(z)$, which is completely determined by the root pattern of $p(\xi)$.

To begin with, it is clear that the integral~\eqref{mainint} can be transformed into an elementary
integral (expressible in terms of rational, trigonometric or hyperbolic functions and their
inverses) if $p(\xi)$ has a multiple root. Indeed, if $p(\xi)$ has a multiple root at infinity
(i.e., if $\xi^4p(1/\xi)$ has a multiple root at the origin) then $\deg p\le 2$, and the
integral~\eqref{mainint} is elementary. Otherwise, if $\xi=\xi_0$ is a multiple (finite) real root
of $p$ the projective transformation $z=(\xi-\xi_0)^{-1}$ transforms $p(\xi)$ into a polynomial
$\hat p(z)$ with a multiple root at infinity, i.e., a polynomial of degree at most two. Finally,
if $p(\xi)$ has a pair of complex conjugate double roots $\xi=\xi_1\pm \iu\xi_2$ then
\[
  p(\xi)=c\big[(\xi-\xi_1)^2+\xi_2^2\big]^2
\]
with $c>0$, and the integral~\eqref{mainint} is again elementary.

In view of the above discussion, we need only consider the case in which all the roots of $p(\xi)$
(real or complex) are simple. In this case~\eqref{mainint} is a genuine elliptic integral, which
can be reduced to its standard Legendre form by the general procedure described, e.g., in
Ref.~\cite{La89}. We present in the appendix a simplified version of this procedure adapted to the
integral~\eqref{mainint}. The conclusion of this analysis is that in all cases the integral
\eqref{mainint} can be expressed in terms of the incomplete elliptic integral of the first kind
\[
  F(\vp,k):=\int_0^\vp\frac{\diff\th}{\sqrt{1-k^2\sin^2\th}}\,,
\]
with $\vp\in(-\pi/2,\pi/2)$ and $0<k<1$.

In the following sections we shall present several examples of algebraic inhomogeneous XX spin
chains, including the Krawtchouk and Lamé chains previously mentioned, for which the
integral~\eqref{mainint} can be computed in closed form by the procedure described above, and thus
the RHS of the asymptotic formula~\eqref{Sasymposc}-\eqref{fnonh} can be readily evaluated. We
shall study the applicability of the latter formula both in the standard situation considered in
the literature of constant $B_n$ and half filling, and also outside this regime. We shall verify
that Eq.~\eqref{Sasymposc} is an excellent approximation for the Rényi entanglement entropy in the
standard situation, but this is not the case for inhomogeneous magnetic fields and/or other
fillings.

\section{The sextic chain}\label{sec.sextic}

As our first example, we shall consider the inhomogeneous XX chain associated with the QES sextic
oscillator potential~\cite{Tu88,Sh89,GKO93}, whose parameters are given (up to irrelevant
constants) by~\cite{FG20}
\begin{equation}\label{JnBn}
  J_n=\sqrt{(n+1)(N-n-1)(\ga+n+1/2)}\,,\qquad
  B_n=-\be n\sqrt{N-1}\,,
\end{equation}
with $\ga=0$ or $\ga>1/2$. We shall start by considering the case $\be=0$, for which the magnetic
field term vanishes identically and the asymptotic approximation~\eqref{Salasy} to the
entanglement entropy should hold. Note that for finite $\ga$ the hopping amplitude $J_n$ is not
symmetric about the chain's midpoint, i.e., $J_{n}\ne J_{N-2-n}$. On the other hand, for
$\ga\to\infty$, or more precisely when $\ga\gg N$, we have
\[
J_n\simeq\sqrt{\ga}\sqrt{(n+1)(N-n-1)}\,,
\]
which is symmetric under $n\mapsto N-n-2$.

To begin with, we write the coefficient $J_n$ as
\begin{equation}\label{Jnsextic}
  J_n=(N-1)^{3/2}\sqrt{\bigg(\frac{x_n}\cL+\frac{1}{N-1}\bigg)\bigg(1-\frac{x_n}\cL\bigg)
    \bigg(\frac{x_n}\cL+\frac{\ga+\frac12}{N-1}\bigg)}\,.
\end{equation}
We shall suppose that the limit
\begin{equation}\label{alimit}
  a:=\lim_{N\to\infty}\frac{\ga+\frac12}{N-1}
\end{equation}
exists. From the restrictions on $\ga$ it follows that $a\ge0$; we shall first analyze the generic
case $a>0$. We can then drop the $1/(N-1)$ term in the first factor under the radical in
Eq.~\eqref{Jnsextic} and take the continuum limit of $J_n$ (after an obvious rescaling) as
\[
  J(x)=\sqrt{p(x/\cL)}\,,\qquad p(\xi)=\xi(1-\xi)(\xi+a)\,,\quad a>0\,.
\]
The integral in Eq.~\eqref{mainint} is most easily computed through the change of variable
$\xi=\cos^2\th$, which yields
\[
  \int_0^s\frac{\diff\xi}{\sqrt{\xi(1-\xi)(\xi+a)}}=2\int_{\arccos\sqrt
    s}^{\frac\pi2}\frac{\diff\th}{\sqrt{a+1-\sin^2\th}}
  =\frac{2}{\sqrt{a+1}}\,\left[K(k)-F\bigl(\arccos\sqrt s,k\bigr)\right]\,,
\]
where
\[
  K(k):=F(\pi/2,k)=\int_0^{\pi/2}\frac{\diff\th}{\sqrt{1-k^2\sin^2\th}}
\]
is the complete elliptic integral of the first kind, and the modulus of the elliptic functions is
\[
  k=(1+a)^{-\frac12}\,.
\]
We thus have (dropping, for the sake of conciseness, the modulus $k$)
\[
  \tx(x)=\frac{2\cL}{\sqrt{a+1}}\,\left[K-F\left(\arccos\sqrt{x/\cL}\,\right)\right]\,,
  \qquad \tcL=\tx(\cL)=\frac{2K\cL}{\sqrt{a+1}}\,,
\]
and hence
\[
  \frac{\tell}{\tcL}=1-\frac{F\bigl(\arccos\sqrt{\la}\,\bigr)}{K}\,,
\]
where we have used the fact that
\[
  \frac{\ell}{\cL}\simeq\frac LN\to\la\,.
\]
Using Eq.~\eqref{fnonh} we finally obtain the following closed-form expression for $f(N,\la)$ when
$a$ is positive:
\begin{equation}\label{fsextic}
  f(N,\la)=\frac{2KN}{\pi\sqrt{a+1}}\,\sqrt{\la(1-\la)(a+\la)}\,\sin\left(\frac{\pi
      F\bigl(\arccos\sqrt{\la}\,\bigr)}{K}\right),\qquad a>0\,.
\end{equation}

Consider next the case in which the limit~\eqref{alimit} vanishes, so that the term $1/(N-1)$ in
the first factor under the radical in Eq.~\eqref{Jnsextic} cannot be neglected. We now write
\[
  p(\xi)=(\xi+\vep_1)(\xi+\vep_2)(1-\xi)\,,
\]
with
\[
  \vep_1:=\frac{\min\left(1,\ga+\frac12\right)}{N-1}<
  \vep_2:=\frac{\max\left(1,\ga+\frac12\right)}{N-1}
\]
small (note that $\vep_1\ne\vep_2$ on account of the conditions $\ga=0$ or $\ga>1/2$). The
integral~\eqref{mainint} is readily computed through the change of variables
$\xi=1-(1+\vep_1)\sin^2\th$. We thus obtain
\[
  \int_0^{s}\frac{\diff\xi}{\sqrt{(\xi+\vep_1)(\xi+\vep_2)(1-\xi)}}
  = \frac2{\sqrt{1+\vep_2}}\,[F_1(0)-F_1(s)],
\]
where
\[
  F_1(s):=F\left(\arcsin\left(\sqrt{\frac{1-s}{1+\vep_1}}\,\right)\right)
\]
and the modulus of the elliptic integral is
\[
  k=\sqrt{\frac{1+\vep_1}{1+\vep_2}}\,<1\,.
\]
Hence
\[
  \tx(x)=\frac{2\cL}{\sqrt{1+\vep_2}}\,[F_1(0)-F_1(x/\cL)],\qquad
  \tcL=\frac{2\cL F_1(0)}{\sqrt{1+\vep_2}}\,,
\]
and therefore
\[
  \tell=\tx(\ell)=\tcL\left[1-\frac{F_1(\la)}{F_1(0)}\right].
\]
Proceeding as before we arrive at the following formula for the function $f(N,\la)$ in
Eq.~\eqref{Sasymposc}:
\[
  f(N,\la)=\frac{2NF_1(0)}{\pi\sqrt{1+\vep_2}}\,\sqrt{p(\la)}\,\sin\left(\frac{\pi F_1(\la)}{F_1(0)}\right).
\]
In the limit $\vep_{1,2}\to0+$ the constant $F_1(0)$ tends to $K(1)=\infty$, while from the identities
\[
  \sn x\underset{k\to1-}\longrightarrow\tanh x\,,\qquad
  F(\vp)=\sn^{-1}(\sin\vp)\underset{k\to1-}\longrightarrow\arctanh(\sin\vp)\,,
\]
it follows that
\[
  F_1(\la)\underset{\vep_1,\vep_2\to0+}\longrightarrow\arctanh(\sqrt{1-\la}\,)\,.
\]
The latter limit is finite for $\la\ne0$, in which case for large $N$ we can write
\[
  \sin\left(\frac{\pi F_1(\la)}{F_1(0)}\right)\simeq\frac{\pi F_1(\la)}{F_1(0)}\,.
\]
Thus when $\la>0$ in the limit~$\vep_{1,2}\to0+$ we have
\begin{equation}\label{Sasympa0}
  f(N,\la)=2N\la\sqrt{1-\la}\arctanh(\sqrt{1-\la}\,)\,,\qquad a=0\,,
\end{equation}
which coincides with the $a\to0+$ limit of Eq.~\eqref{fsextic}.

It is also straightforward to compute the $a\to\infty$ limit of the function $f(N,\la)$
in~\eqref{fsextic}. Indeed, in this limit the modulus $k=(1+a)^{-\frac12}$ tends to zero, so that
$K\to\pi/2$, $F(\vp)\to\vp$, and therefore
\[
  \sin\left(\frac{\pi}K\,F\left(\arccos\sqrt{\la}\,\right)\right)
  \to\sin\left(2\arccos\sqrt{\la}\,\right)=2\sqrt{\la(1-\la)}\,.
\]
We thus obtain the asymptotic formula
\begin{equation}\label{flaainf}
 \lim_{a\to\infty}f(N,\la)=2N\la(1-\la)\,.
\end{equation}
Note that the right-hand side of the latter equation is invariant under $\la\mapsto 1-\la$, i.e.,
$L\mapsto N-L$. This is due to the fact that $a\gg 1$ implies that $\ga\gg N$, and hence $J_n$ is
approximately symmetric about $n=N/2-1$. In such symmetric chains the entanglement entropy is
necessarily invariant under $L\mapsto N-L$, since
\begin{equation}\label{Ssymm}
  S_\al\bigl[\{0,\dots,L-1\}\bigr]=S_\al\bigl[\{L,\dots,N-1\}\bigr]
  =S_\al\bigl[\{0,\dots,N-L-1\}\bigr]\,,
\end{equation}
where the first equality follows from Schmidt's decomposition and the second one is due to the
chain's symmetry about its midpoint. On the other hand, for finite $a$ the sextic chain is not
symmetric about its midpoint, and thus neither its entanglement entropy nor the asymptotic
approximation~\eqref{Sasymposc} thereof are invariant under $\la\mapsto 1-\la$.

\begin{figure}[t]
  \includegraphics[height=.33\textwidth]{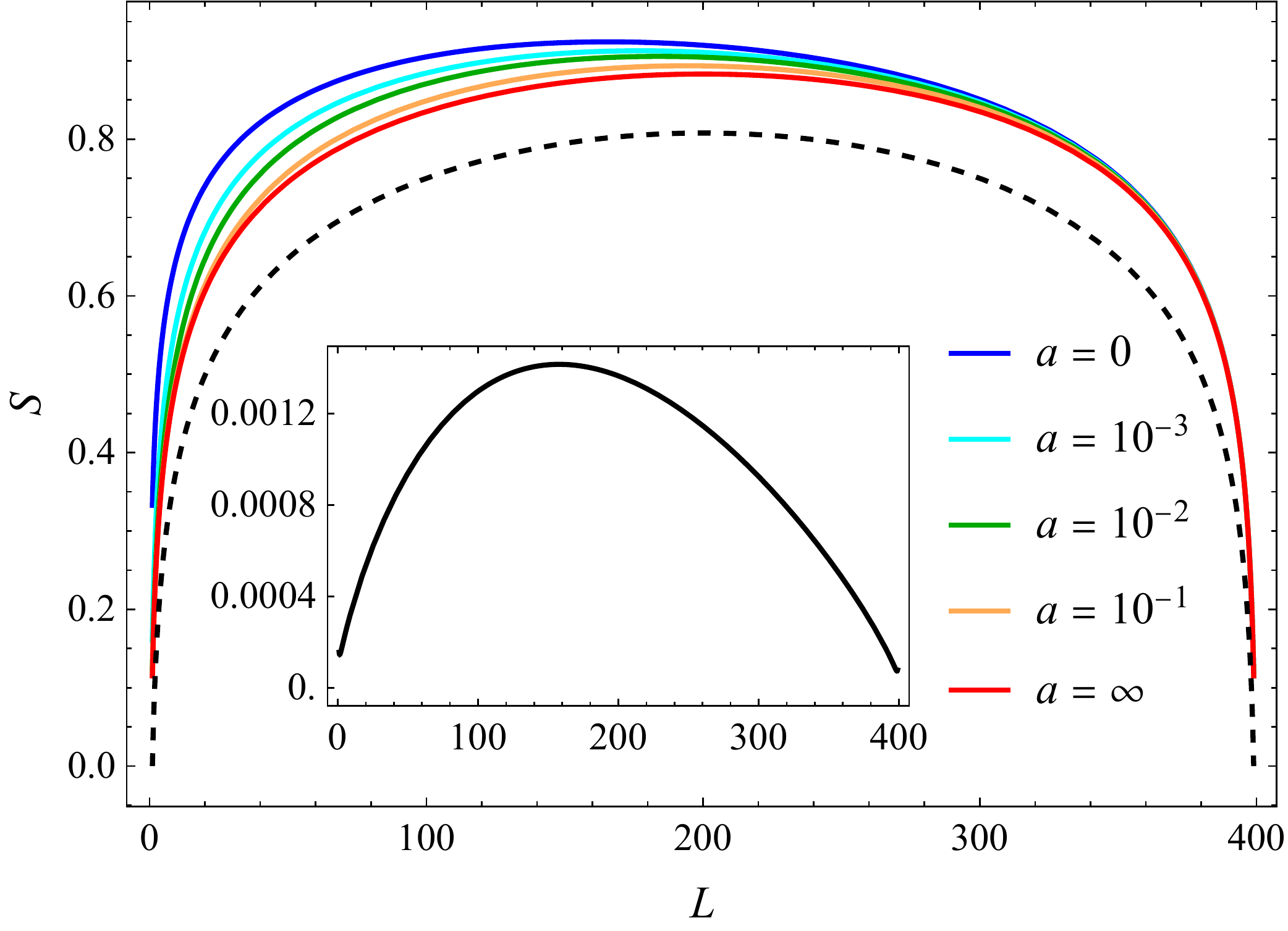}\hfill
  \includegraphics[height=.33\textwidth]{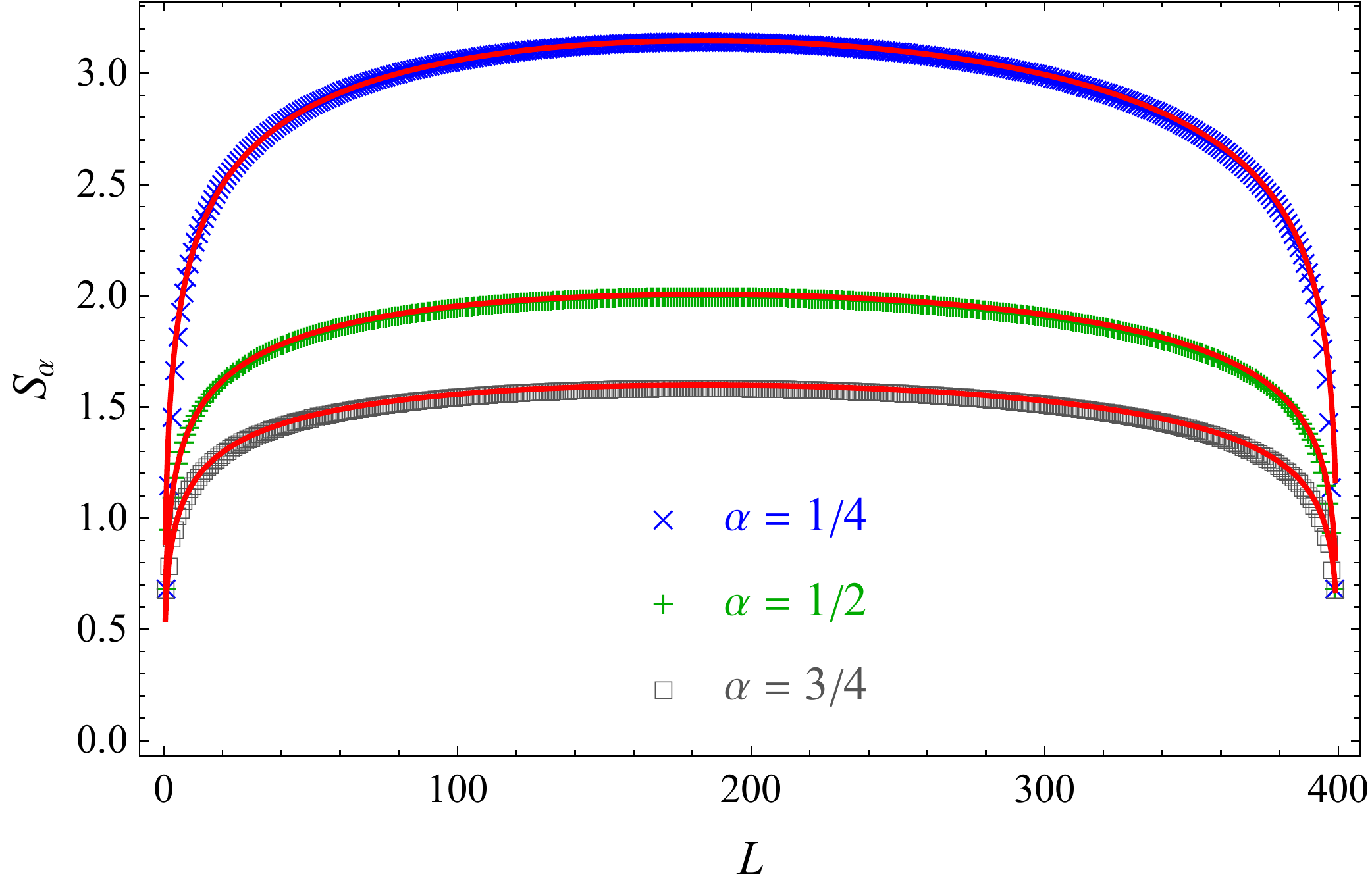}
  \caption{Left: leading term in the asymptotic approximation~\eqref{Sasymposc} to the von Neumann
    entanglement entropy of the sextic chain with $N=400$ spins for
    $a=0,10^{-3},10^{-2},10^{-1},\infty$ (the dashed black line represents the analogous quantity
    for the homogeneous XX chain). Inset: relative error between the $a=1$ and $a=\infty$
    approximations. Right: Rényi entanglement entropy of the sextic chain with $N=400$ spins and
    $a=10^{-2}$ for $\al=1/4,1/2,3/4$, compared to its asymptotic approximation~\eqref{Sasymposc}
    (solid red lines).}
  \label{fig.f-sextic}
\end{figure}

From the explicit expressions~\eqref{fsextic}-\eqref{Sasympa0} of the function $f(N,\la)$ we can
easily deduce the behavior of the leading term in the asymptotic approximation~\eqref{Sasymposc}.
Since this term depends trivially on $\al$, in Fig.~\ref{fig.f-sextic} (left) we present only a
plot of the leading order approximation~$S_{\text{app}}(N,\la):=(1/6)\log f(N,\la)$ to the von
Neumann entanglement entropy of the sextic chain for $N=400$ spins and several values of the
parameter $a$, including the limiting cases $a=0$ and $a=\infty$. It is apparent that
$S_{\text{app}}(N,\la)$ decreases monotonically with $a$, and that the graph of
$S_{\text{app}}(N,\la)$ approaches that of its $a\to\infty$ limit~\eqref{flaainf} even for values of
$a$ as low as $10^{-1}$. In fact, for $a=1$ the relative error between $S_{\text{app}}(N,\la)$ and
Eq.~\eqref{flaainf} is less that $1.4\cdot 10^{-3}$ (cf.~inset of Fig.~\ref{fig.f-sextic} (left)),
so that both graphs are virtually indistinguishable. On the other hand, the approach of the graph
of $S_{\text{app}}(N,\la)$ to its limit~\eqref{Sasympa0} as $a\to0+$ is much slower, particularly
for $L<N/2$ (see, e.g., the $a=10^{-3}$ graph in Fig.~\ref{fig.f-sextic} (left)).

\begin{figure}[t]
  \includegraphics[height=.32\textwidth]{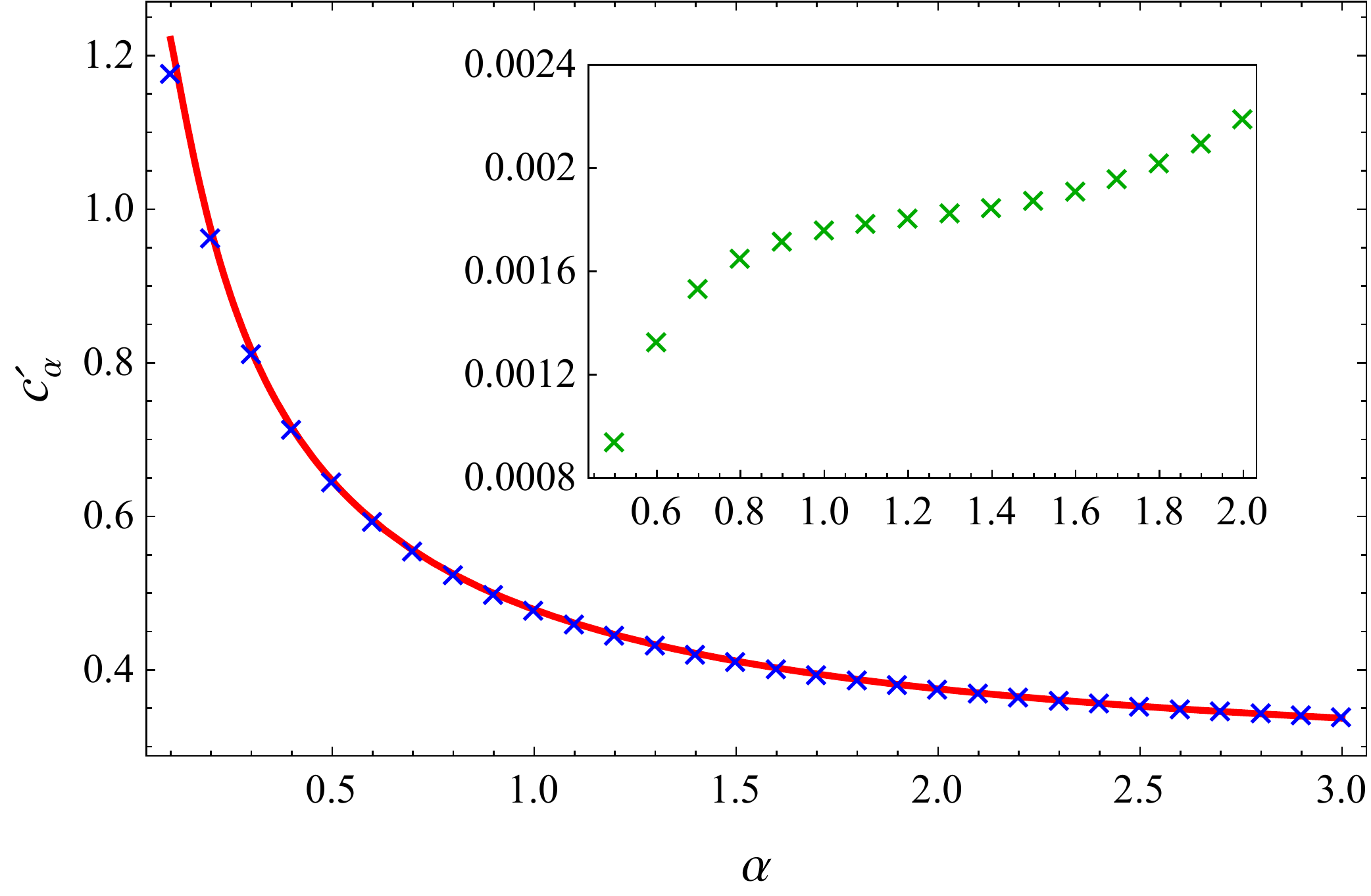}\hfill
   \includegraphics[height=.32\textwidth]{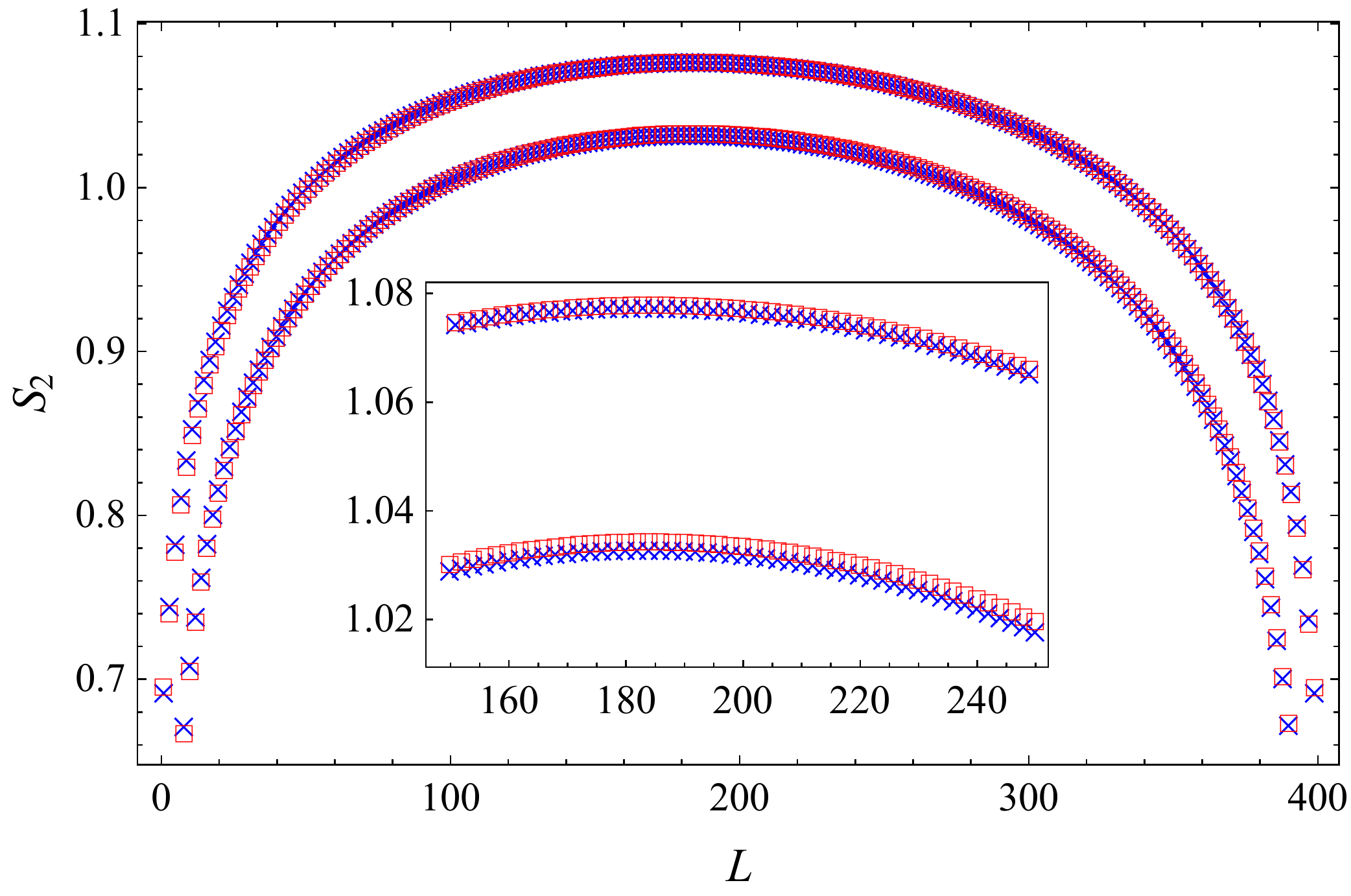}
   \caption{Left: constant term $c'_\al$ in the asymptotic
     approximation~\eqref{Sasymposc}-\eqref{fnonh} to the Rényi entanglement entropy $S_\al$ of
     the sextic chain with $a=10^{-2}$ and $N=400$ spins for $\al=i/10$ and $1\le i\le 30$ (blue
     crosses) compared to the corresponding constant $c'_\al$ for the homogeneous XX chain in
     Eq.~\eqref{cpal} (solid red line). The inset shows the difference between $c_\al'$ and its
     counterpart~\eqref{cpal} for the homogeneous chain in the range $1/2\le\al\le 2$ at intervals
     of $1/10$. Right: Rényi entanglement entropy $S_2$ for the sextic chain with $a=10^{-2}$ and
     $N=400$ spins at half filling and zero magnetic field (blue crosses) compared to its
     asymptotic approximation~\eqref{Sasymposc} (red squares). The inset shows a blow up of the
     range $150\le L\le 250$ in the latter plot.}
  \label{fig.S1434}
\end{figure}

Our numerical simulations indicate that the asymptotic formula~\eqref{Sasymposc}-\eqref{fnonh}
does indeed provide an excellent approximation to the Rényi entanglement of the sextic chain for
$N\gg1$ in the absence of a magnetic field and at half filling. For $\al<1$ this is illustrated by
Fig.~\ref{fig.f-sextic} (right), where we present the case of $N=400$ spins for $a=10^{-2}$ (for
which the hopping amplitude is neither homogeneous nor approximately symmetric about the midpoint)
and $\al=1/4,1/2,3/4$. Of course, in order to compare the asymptotic
formula~\eqref{Sasymposc}-\eqref{fnonh} with the exact (numerically computed) value of the
entanglement entropy $S_\al$ it is necessary to first determine the constant part $c'_\al$ in the
former equation. We have simply estimated $c'_\al$ as the average value of the difference between
$S_\al$ and the leading term of its asymptotic approximation~\eqref{Sasymposc}. Surprisingly, this
value of $c'_\al$ coincides to a remarkable accuracy with the corresponding one for the
homogeneous XX chain given by Eq.~\eqref{cpal} (see, e.g., Fig.~\ref{fig.S1434} (left) for
$a=10^{-2}$). In fact, we have checked that this is also the case for several other values of the
parameter $a$.

As mentioned in the previous section, in the homogeneous XX chain the $o(1)$ term in the
asymptotic formula~\eqref{SCFT} for the Rényi entanglement entropy with
parameter~$\al\ge\nobreak1$ is oscillatory and of order $N^{{-1/\al}}$
(cf.~Eqs.~\eqref{Sasymphom}-\eqref{mutheor}). In particular, at half filling this term features
parity oscillations with amplitude roughly proportional to $[(N/\pi)\sin(\pi L/N)]^{-1/\al}$. We
have checked that the behavior of the Rényi entanglement entropy of the sextic chain with
parameter $\al\ge1$ at half-filling and zero magnetic field is very similar, and in particular
that its parity oscillations are reproduced with great accuracy by
Eqs.~\eqref{Sasymposc}-\eqref{fnonh}. This can be seen, for instance, in Fig.~\ref{fig.S1434}
(right), where we compare the Rényi entanglement entropy with parameter $\al=2$ for $a=10^{-2}$
and $N=400$ spins with its asymptotic approximation~\eqref{Sasymposc}. Remarkably, in all the
cases we have analyzed the values of the parameters $c_\al'$ and $\mu_\al$ are very close to the
corresponding ones for the homogeneous model, given by Eqs.~\eqref{cpal} and~\eqref{mutheor}. This
suggests ---as shall be further corroborated by the analysis of the Krawtchouk chain in the next
section--- that at half filling and in a vanishing (or constant) magnetic field the sextic chain
is in the same universality class as the homogeneous XX chain.

\begin{figure}[t]
  \includegraphics[height=.32\textwidth]{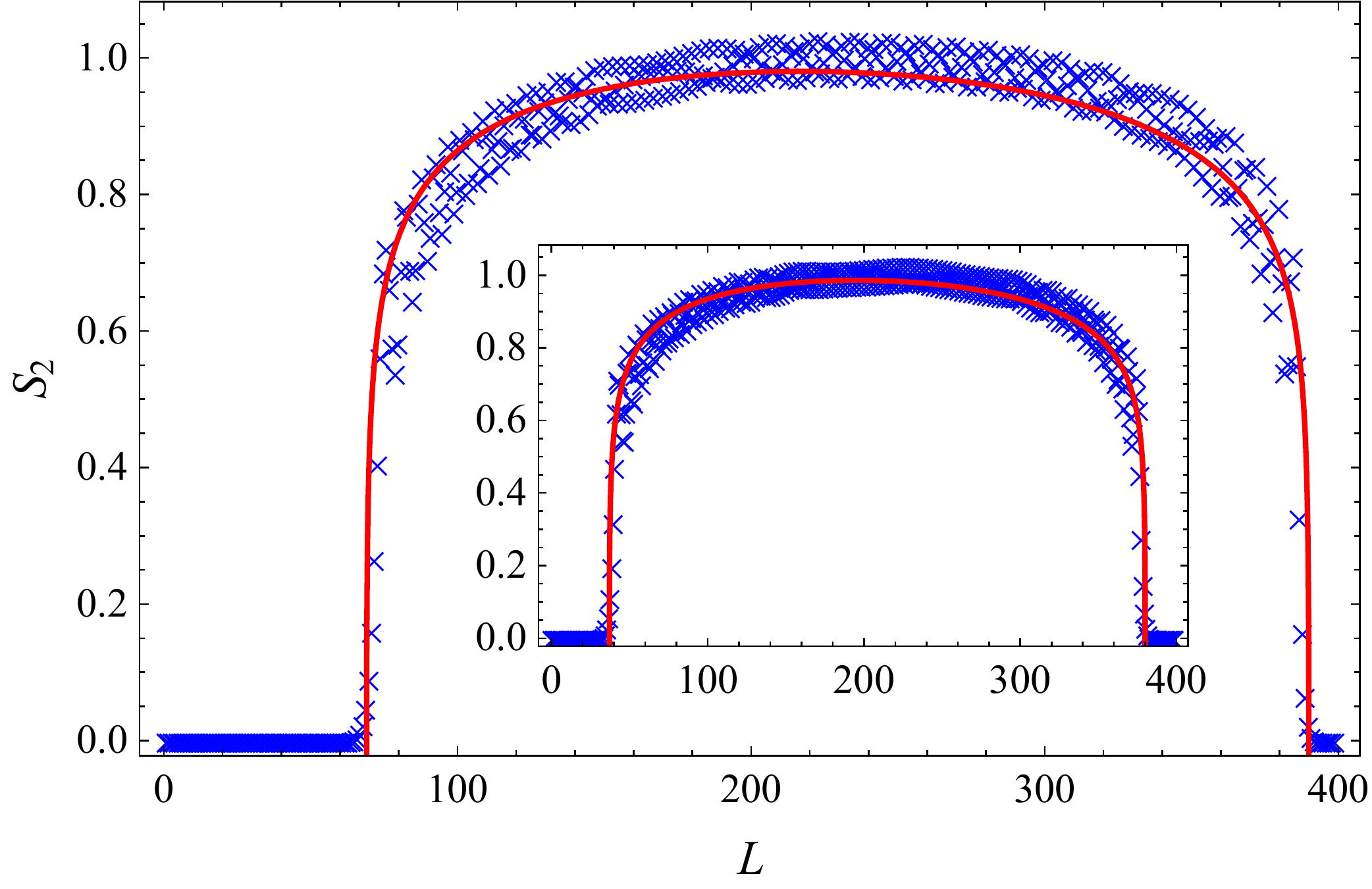}\hfill
  \includegraphics[height=.32\textwidth]{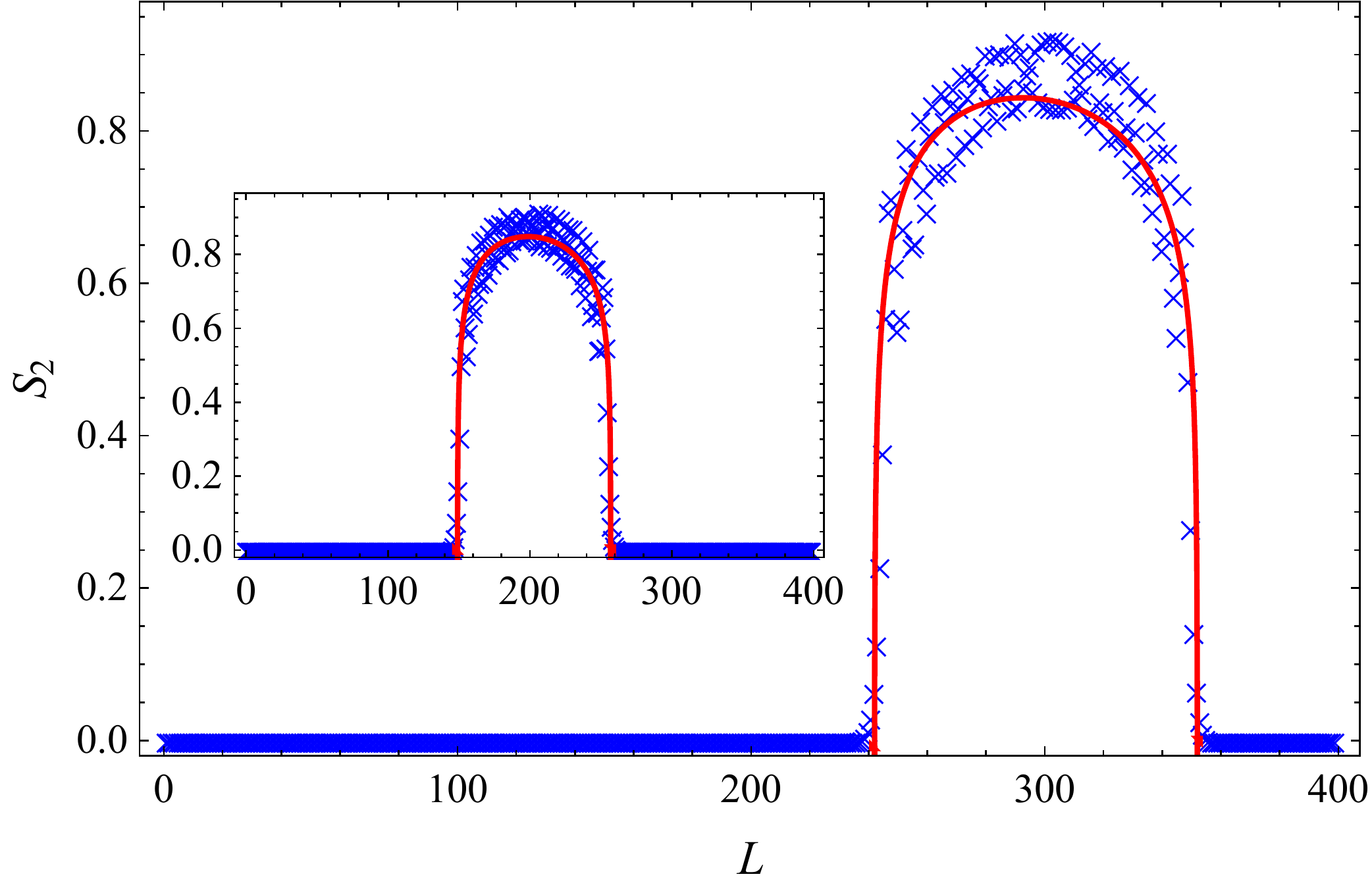}
  \caption{Left: Rényi entanglement entropy $S_2$ for the sextic chain with $N=400$ spins,
    $a=10^{-2}$, $\be=0$, and $k_F=\pi/4$ (blue crosses) compared to its rough
    approximation~\eqref{Salbe}-\eqref{fapp} (solid red line). The inset shows the same plot for
    $a=1$. Right: Analogous plot for $a=10^{-2}$, $\be=5$ and $k_F=\pi/4$ (main plot) or
    $k_F=\pi/2$ (inset).}
  \label{fig.badfits}
\end{figure}

The situation is markedly different in the presence of an inhomogeneous magnetic field and/or at
arbitrary fillings. Indeed, our numerical calculations clearly indicate that in these cases the
behavior of the Rényi entanglement entropy is not well described even to leading order by the
conformal analogue of Eq.~\eqref{SCFTkF}-\eqref{fHdef}, namely
\begin{equation}
  \label{Salbe}
  S_\al(N,\la)=\frac1{12}\,\big(1+\al^{-1}\big)\log f(N,\la)+c_\al'+o(1)\,,
\end{equation}
with
\begin{align}\label{fkF}
  f(N,\la)&=\frac{\tcL}{\pi\De \wt x}\sin\left(\frac{\pi\tell}{\tcL}\right)\sin k_F\nonumber\\
  &=\frac{2KN}{\pi\sqrt{a+1}}\,\sqrt{\la(1-\la)(a+\la)}\,\sin\left(\frac{\pi
      F\bigl(\arccos\sqrt{\la}\,\bigr)}{K}\right)\sin k_F\,.
\end{align}
Consider, to begin with, the case $\be=0$ and $k_F\ne\pi/2$, illustrated in Fig.~\ref{fig.badfits}
(left) for the Fermi momentum $k_F=\pi/4$ and $a=10^{-2}$ or $a=1$. The fact that $k_F\ne\pi/2$ is
seen to have two main effects. In the first place, $S_\al$ is now virtually zero for small $L$ and
$N-L$ (for instance, if $N=400$ and $a=10^{-2}$ then $S_2$ is less than $10^{-3}$ for
$1\le L\le64$ and $393\le L\le 399$, while for $a=1$ we have $S_2<10^{-3}$ if $1\le L\le31$ and
$385\le L\le 399$). It is then clear that the entanglement entropy cannot be well approximated in
this case by the concave function in the RHS of Eqs.~\eqref{Salbe}-\eqref{fkF}. Moreover, for
$\al\ge1$ the parity oscillations of $S_\al$ are much less regular than in the case of half
filling, and their amplitude is not well reproduced by a simple formula like~\eqref{Sasymposc}
(see, e.g., the main plot in Fig.~\ref{fig.badfits} (left) for the case $a=10^{-2}$, $\be=0$,
$k_F=\pi/4$, $N=400$ and $\al=2$). On the other hand, a rough approximation capturing only the
average variation of $S_\al$ with $L$ can be obtained by restricting ourselves to the interval
$[L_1+1,L_2-1]$ in which $S_\al$ differs significantly from zero, and replacing accordingly $L$
and $N$ respectively by $L-L_1$ and $L_2-L_1$. With these changes Eq.~\eqref{fkF} becomes
  \begin{equation}
    \label{fapp}
    f(N,\la)=\frac{2K(L_2-L_1)}{\pi\sqrt{a+1}}\,
    \sqrt{\laeff(1-\laeff)(a+\laeff)}\,\sin\left(\frac{\pi
        F\bigl(\arccos\sqrt{\laeff}\,\bigr)}{K}\right)\sin k_F\,,
  \end{equation}
  with
  \begin{equation}
    \laeff=\frac{L-L_1}{L_2-L_1}=\frac{N\la-L_1}{L_2-L_1}\,.
    \label{laeff}
  \end{equation}
The rough approximation~\eqref{Salbe}-\eqref{fapp} is represented by a red line in
Fig.~\ref{fig.badfits}.

The situation is qualitatively similar in the presence of an external magnetic field given by
Eq.~\eqref{JnBn} with $\be\ne0$, even in the case of half filling; see, e.g.,
Fig.~\ref{fig.badfits} (right). More precisely, as seen in the latter figure, the length of the
intervals in which $S_\al$ is practically zero increases significantly with $\be$. Moreover, when
$k_F$ differs from $\pi/2$ the interval over which $S_\al$ is appreciably different from zero is
translated by an amount depending on $\be$. As before, the pattern of the parity oscillations is
much more involved than in the case $B_n=0$, $k_F=\pi/2$ analyzed earlier, although the average
variation of $S_\al$ with $L$ is still reproduced to a certain extent by the heuristic
formula~\eqref{Salbe}-\eqref{fapp}.

Note, finally, that the fact that the entropy of the block~$\{0,\dots,L-1\}$ is negligibly small
for $L\le L_1$ and $L\ge L_2$ clearly indicates that the first $L_1$ and last $N-L_2$ spins are
approximately in a product state, so that the chain's entanglement is almost entirely concentrated
in the central block~$\{L_1,\dots,L_2-1\}$. It would certainly be of interest to understand how
exactly this phenomenon arises as the external magnetic field is turned on, or the standard
filling $M/N=1/2$ is varied.

\section{The Krawtchouk chain}\label{sec.Krawtchouk}

\subsection{Definition and entanglement entropy}
The Krawtchouk chain was introduced in Ref.~\cite{CNV19} in connection with the family of discrete
Krawtchouk polynomials. More precisely, the Krawtchouk polynomial $K_n(x;q,m)$ is defined as
\begin{equation}\label{KrawP}
  K_n(x;q,m)={}_2F_1(-n,-x;-m;1/q)\,,\qquad n=0,\dots,m\,,
\end{equation}
where $0<q<1$ and $m$ is a nonnegative integer (see, e.g.,~\cite{KLS10}). Here ${}_2F_1$ denotes
the standard hypergeometric function
\[
  {}_2F_1(a,b;c;z):=\sum_{k=0}^\infty\frac{(a)_k(b)_k}{(c)_k}\,\frac{z^k}{k!}\,,
\]
where $(a)_k:=a(a+1)\cdots(a+k-1)$ is the usual (ascending) Pochhammer symbol. These polynomials
satisfy the recursion relation
\[
  A_nK_{n+1}(x)=(B_n-x)K_n(x)-C_nK_{n-1}(x)
\]
with
\[
  A_n=q(m-n)\,,\qquad B_n=n(1-q)+q(m-n)\,,\qquad C_n=n(1-q)\,.
\]
The corresponding monic polynomials are given by
\[
  P_n(x)=\prod_{k=0}^{n-1}(-A_k)\cdot K_n(x;q,m)=(-q)^n(m-n+1)_nK_n(x;q,m)\,,
\]
and satisfy the normalized recursion relation
\[
  P_{n+1}(x)=(x-B_n)P_n(x)-a_nP_{n-1}(x)
\]
with $B_n$ as before and
\[
  a_n=A_{n-1}C_n=q(1-q)n(m-n+1)>0\,,\qquad 1\le n\le m\,.
\]
The polynomial $K_{m+1}(x;q,m)$ cannot be defined through Eq.~\eqref{KrawP}, since $(-m)_{m+1}=0$.
On the other hand, we can use the previous recursion relation with $n=m$ to define $P_{m+1}$ by
\[
  P_{m+1}(x)=\big[x-m(1-q)\big]P_m(x)-mq(1-q)P_{m-1}(x)\,.
\]
Using the previous formula and the recursion relation, after a long but straightforward
calculation we obtain
\[
  P_{m+1}(x)=\prod_{k=0}^m(x-k)\,.
\]
In view of the above, we define the Krawtchouk chain\footnote{This definition differs from the one
  in Ref.~\cite{CNV19} only in the sign of the magnetic field.} through the polynomials $P_n(x)$
with $m=N-1$. In other words, the chain's parameters $J_n$ and $B_n$ are given by
\begin{align}\label{JKraw}
  J_n&=\sqrt{a_{n+1}}=\sqrt{q(1-q)(n+1)(N-n-1)}\,,\\
  B_n&=n(1-q)+q(N-n-1)=q(N-1)+(1-2q)n\,.
  \label{BKraw}
\end{align}
In particular, the magnetic field is constant if and only if $q=1/2$. Using the definition of the
Krawtchouk polynomials we readily obtain the explicit formula
\[
  P_n(E)=(N-n)_n\sum_{k=0}^n\frac{\binom nk}{\binom{N-1}k}\,
  \frac{(-q)^{n-k}}{k!}\,E(E-1)\cdots(E-k+1)\,,\qquad n=0,\dots,N-1\,,
\]
with $P_N(E)=E(E-1)\cdots(E-N+1)$. Thus in this case the single-particle energies $\vep_k$ are
simply the integers $0,1,\dots,N-1$. This makes it possible to express the matrix elements
$\Phi_{nk}$ in closed form using Eqs.~\eqref{phiP}-\eqref{gaw}. Indeed, it is readily found that
\[
  \ga_n=q^n(1-q)^nn!(N-n)_n\,,\qquad w_k=\binom{N-1}kq^k(1-q)^{N-1-k}\,,
\]
and therefore

\begin{align}
  \Phi_{nk}&=\frac1{n!}\sqrt{\frac{\binom{N-1}k}{\binom{N-1}n}}\,q^{\frac12(k-n)}
             (1-q)^{\frac12(N-k-n-1)}\,P_n(k)\nonumber\\
  &=(-1)^n\sqrt{\binom{N-1}k\binom{N-1}n}\,q^{\frac12(k+n)}
    (1-q)^{\frac12(N-k-n-1)}\,K_n(k;q,N-1)\,.
    \label{PhiKraw}
\end{align}

\begin{figure}[t]
  \includegraphics[height=.32\textwidth]{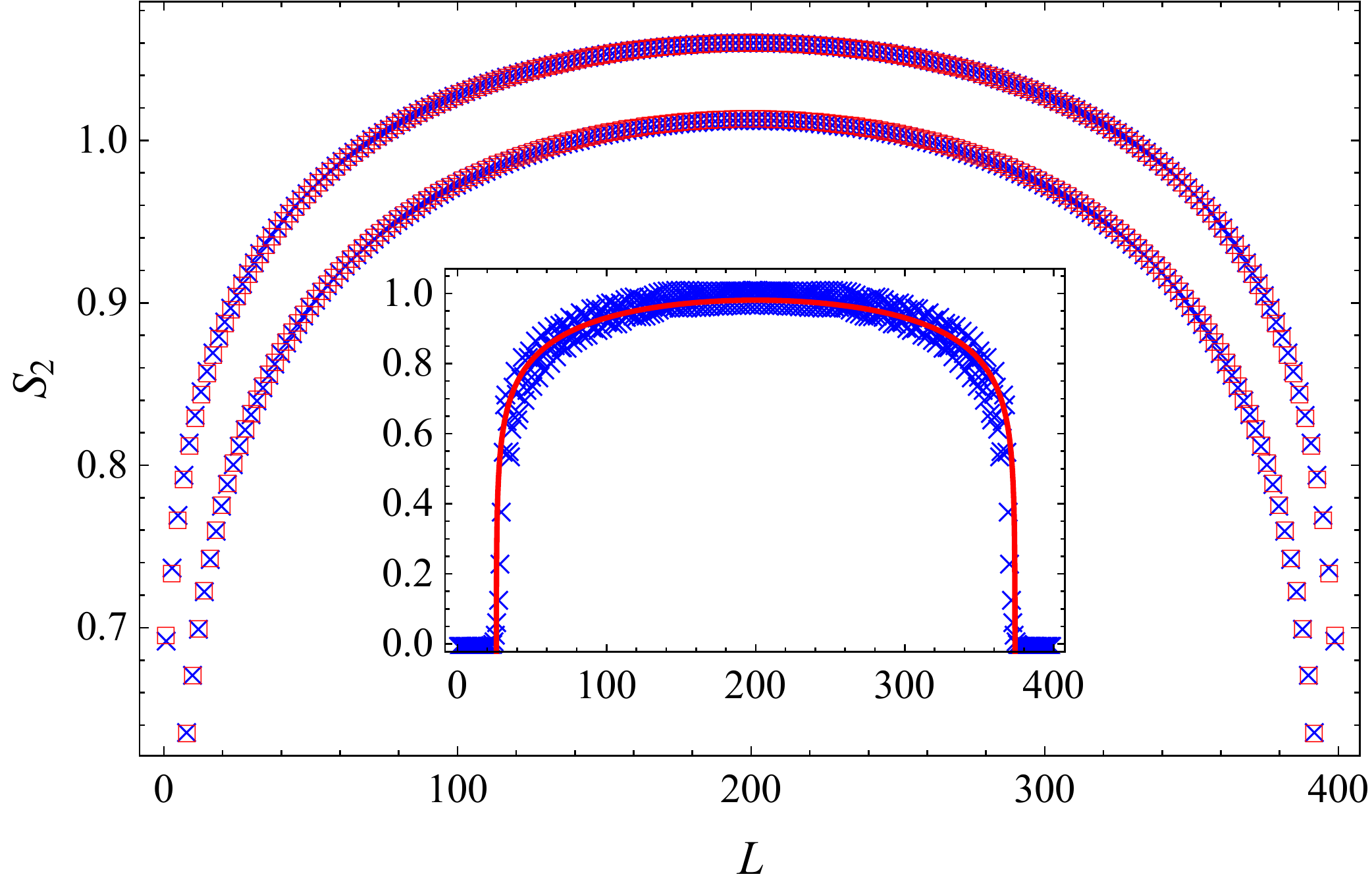}\hfill
  \includegraphics[height=.32\textwidth]{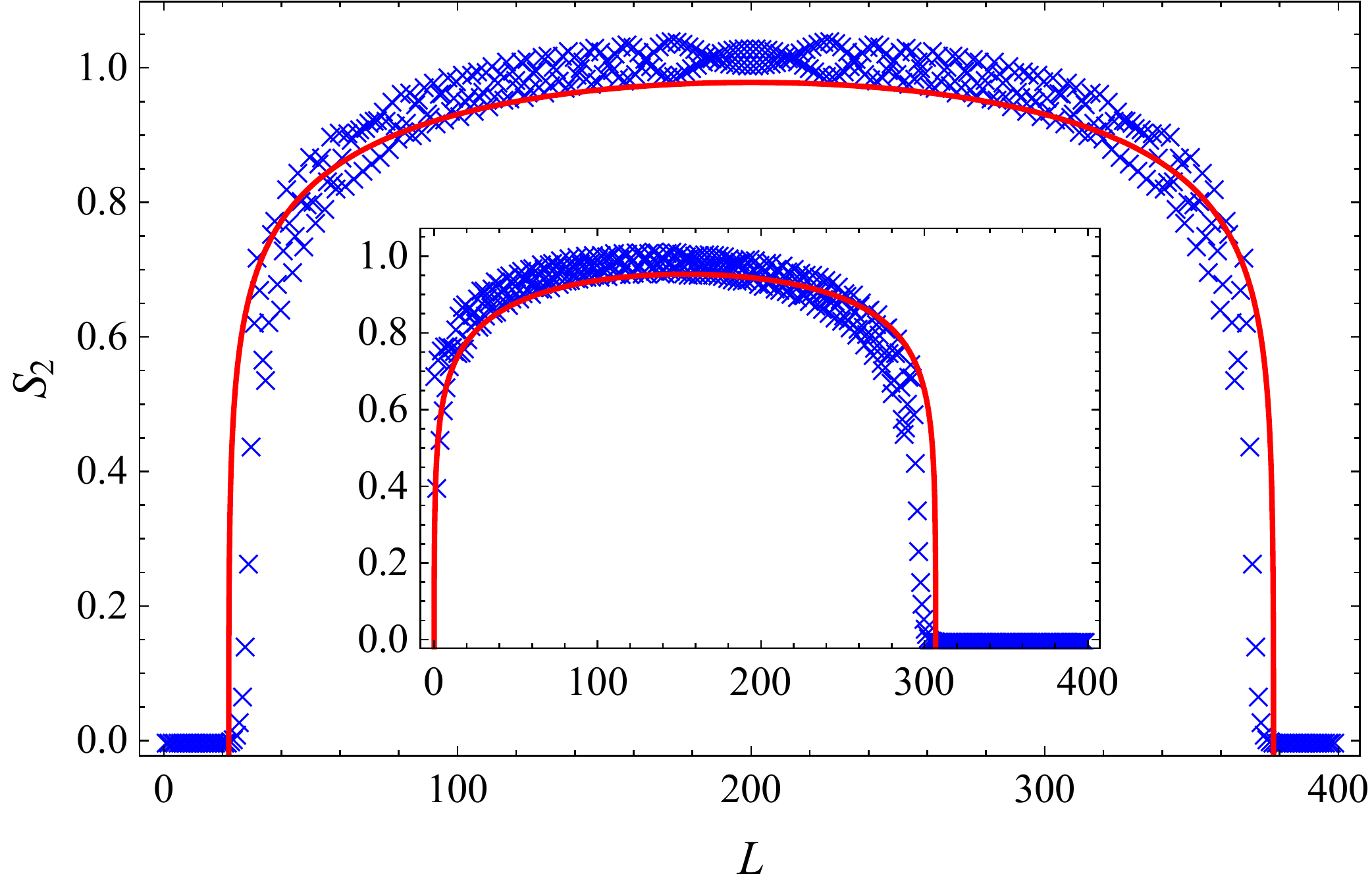}
  \caption{Left: Rényi entanglement entropy $S_2$ for the Krawtchouk chain with $q=1/2$ and
    $N=400$ spins at half filling (blue crosses) compared to its asymptotic
    approximation~\eqref{Sasymposc}-\eqref{fKr} (red squares). The inset shows a similar plot for
    $k_F=\pi/4$, compared to its heuristic approximation~\eqref{Salbe}-\eqref{fL1L2}. Right:
    analogous plot for $q=1/4$ at half filling (main plot) and for $k_F=\pi/4$ (inset).}
  \label{fig.SKr}
\end{figure}

Dropping the irrelevant overall factor $[q(1-q)]^{1/2}(N-1)$ in Eq.~\eqref{JKraw} we easily obtain
the following formula for the continuum limit of $J_n$:
\[
  J(x)=\sqrt{p(x/\cL)}\,,\qquad p(\xi):=\sqrt{\xi(1-\xi)}\,.
\]
We thus have
\begin{equation}\label{txtLKr}
  \tx(x)=\cL\int_0^{x/\cL}\frac{d\xi}{\sqrt{\xi(1-\xi)}}=2\cL\arcsin\bigl({
      \textstyle\sqrt{x/\cL}}\,\bigr)\,,\qquad
  \tcL=\pi\cL\,,
\end{equation}
and hence the function $f(N,\la)$ in Eq.~\eqref{fnonh} is simply given by
\begin{equation}\label{fKr}
  f(N,\la)=N\sqrt{\la(1-\la)}\,\sin(2\arcsin\sqrt\la)=2N\la(1-\la)\,.
\end{equation}
As expected, this result coincides with the $a\to\infty$ limit of the analogous function for the
sextic chain. In other words, the asymptotic behavior of the entanglement entropy of the
Krawtchouk chain with $q=1/2$ at half filling, given by Eqs.~\eqref{Sasymposc}-\eqref{fKr}, should
be the same as for the sextic chain with $\be=0$ and $a\gg1$. This is shown in Fig.~\ref{fig.SKr}
(left) for $N=400$ spins. For the same reason, at arbitrary fillings the heuristic asymptotic
approximation to the entanglement entropy is given by Eq.~\eqref{Salbe} with
\begin{equation}\label{fL1L2}
  f(N,\la)=\frac{2(N\la-L_1)(L_2-N\la)}{L_2-L_1}\,\sin k_F\,,
\end{equation}
where $[L_1+1,L_2-1]$ is the interval over which $S_\al$ is appreciably nonzero; see, e.g., the
inset of Fig.~\ref{fig.SKr} (left). Note also that when $q=1/2$ we have $J_{n}=J_{N-n-2}$ and
$B_n$ is constant, so that in this case the entanglement entropy is invariant under
$L\mapsto N-L$, i.e., satisfies Eq.~\eqref{Ssymm}, for all fillings.

On the other hand, for $q\ne1/2$ the magnetic field strength $B_n$ is non-uniform, so that the
entanglement entropy behaves much the same as for the sextic chain with $a\gg1$ and $\be\ne0$;
see, e.g., Fig.~\ref{fig.SKr} (right) for the case $q=1/4$ and $N=400$. The main difference, as
seen in the right inset of the latter figure, is that in this case the entropy is close to zero
only on an interval of the form $[L_2,N]$.
%
% Changed
%
\begin{remark}
  The value of the subleading (constant) term $c_\al'$ in the asymptotic
  formula~\eqref{Sasymposc}-\eqref{fKr} for the entanglement entropy of the Krawtchouk chain with
  $q=1/2$ at half filling is remarkably close to its counterpart $c_{\al,\text{hom}}'$ for the
  homogeneous chain (cf.~Eq.~\eqref{cpal}), even more so than in the case of the sextic chain
  discussed above. For instance, the difference $|c_\al'-c_{\al,\text{hom}}'|$ is of the order of
  $10^{-3}$ or less for $1/2\le\al\le2$. In fact, the values of $c_\al'$ in the latter range were
  obtained as in the previous example by taking the average of the differences
  $S_{\al}-S_{\al,\text{app}}$ for all values of the block size $L=1,\dots,N-1$, where
  $S_{\al,\text{app}}$ denotes the leading term in Eq.~\eqref{Sasymposc}. The behavior of the
  latter differences, however, clearly suggests that in this case a more accurate estimate for
  $c_\al'$ is given by the average of $S_{\al}-S_{\al,\text{app}}$ for the central block sizes
  $L=N/2$ and $L=N/2-1$, or more simply by
  \begin{equation}\label{cpalnew}
    c_{\al'}=\frac12\left(S_\al\Big|_{L=N/2-1}+S_\al\Big|_{L=N/2}\right)-S_{\al,\text{app}}\Big|_{L=N/2}\,;
  \end{equation}
  see, e.g., Fig.~\ref{fig.Sdiffs} (left) for $\al=1$. The previous equation actually yields a
  value of $c_\al'$ much closer to $c_{\al,\text{hom}}'$ than the average of the differences
  $S_\al-S_{\al,\text{app}}$ for all block sizes. For instance, for $\al=1$ the difference between
  $c_{1,\text{hom}}'$ and the value of $c_1'$ computed from the previous formula is
  $1.1\cdot 10^{-6}$, compared to $3.6\cdot 10^{-4}$ when $c_1'$ is estimated by the average of
  $S-S_{\text{app}}$ over all block sizes.
  \begin{figure}[t]
    \includegraphics[height=.3\textwidth]{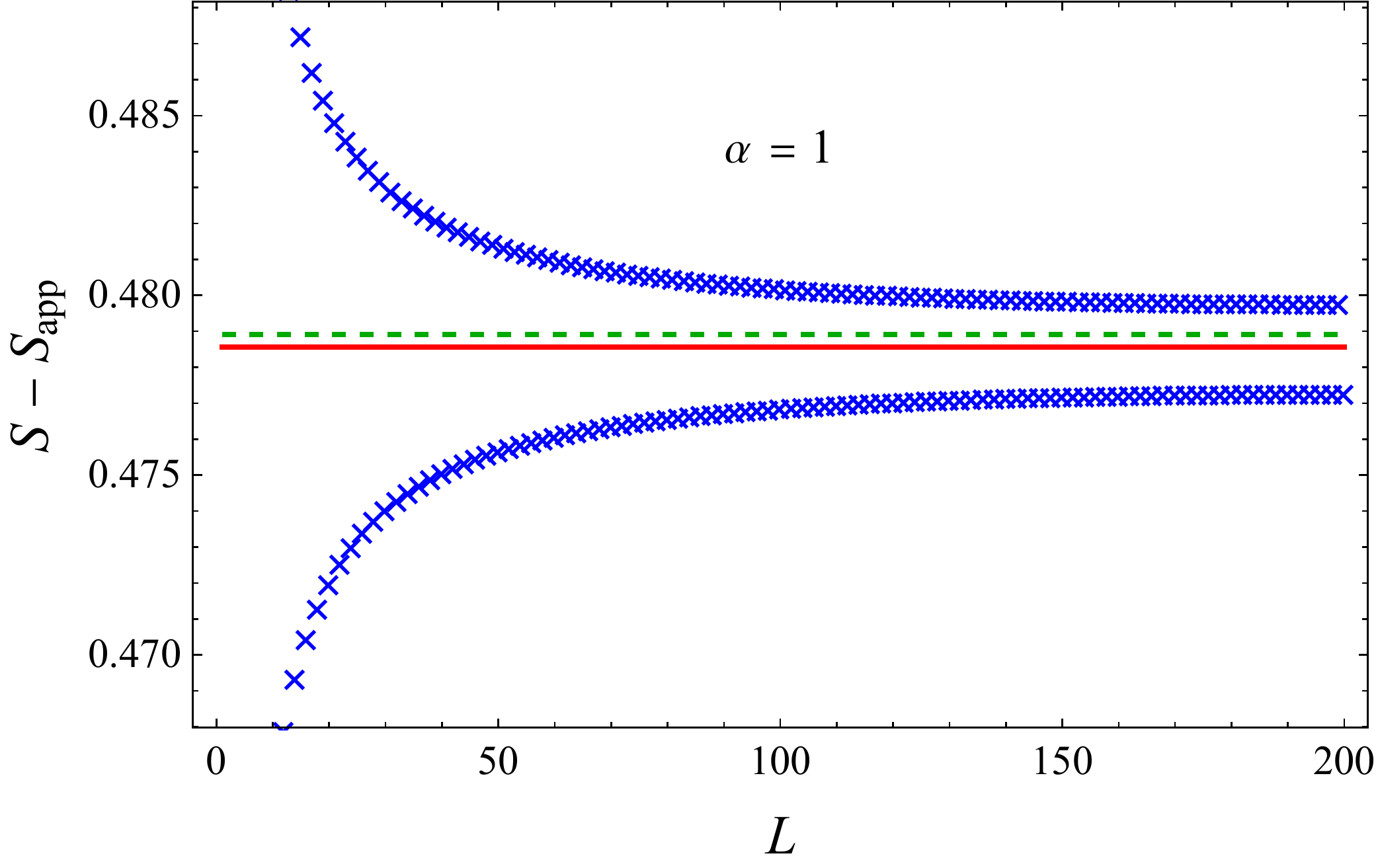}\hfill
    \includegraphics[height=.3\textwidth]{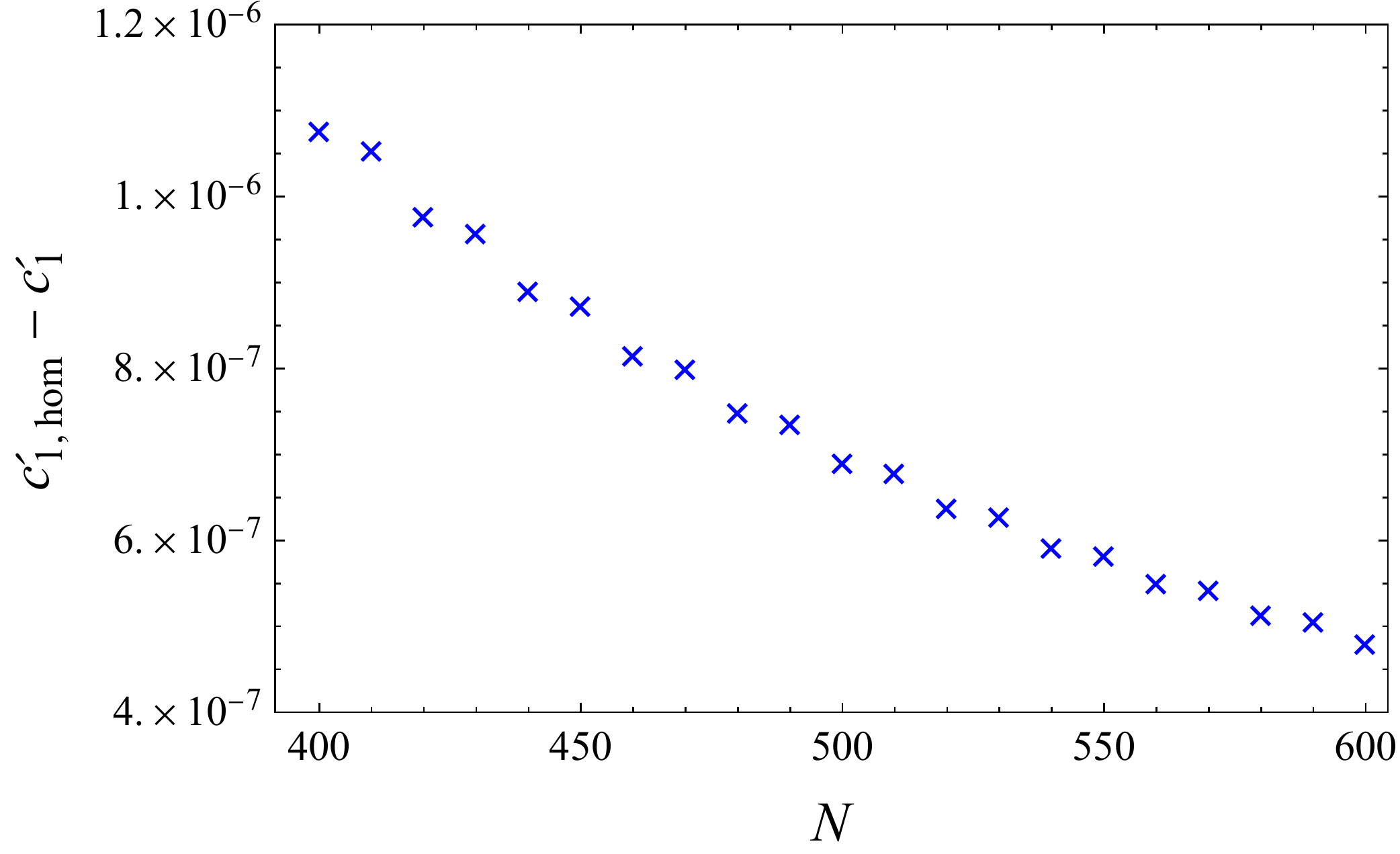}
    \caption{Left: difference between the von Neumann entanglement entropy $S$ of the Krawtchouk
      chain with $q=1/2$, $N=400$, $k_F=\pi/2$ and its leading asymptotic approximation
      $S_{\text{app}}:=(1/6)\log f(N,\la)$ as a function of the block size $L$. (Only the range
      $L=1,\dots,N/2$ has been represented, since in this case $S_\al$ is symmetric under
      $L\mapsto N-L$.) The height of the red (resp.~green dashed) horizontal line is the value of
      $c_1'$ computed from Eq.~\eqref{cpalnew} (resp.~as the average of the differences
      $S-S_{\text{app}}$ for all block sizes). Right: Difference $c_{1,\text{hom}}'-c_1'$ for
      $N=400,\dots,600$ in increments of 10.}
    \label{fig.Sdiffs}
  \end{figure}
  
  We have also studied how the difference $c_{\al,\text{hom}}-c_\al'$ varies as the number of
  spins increases from $N=400$ to $N=600$ (in increments of $10$) for several values of $\al$,
  where in view of the previous remark we have used Eq.~\eqref{cpalnew} to estimate $c_\al'$. As
  is apparent from Fig.~\ref{fig.Sdiffs} (right) for the case $\al=1$, the absolute value of this
  difference steadily decreases with $N$. We thus conjecture that for the Krawtchouk chain with
  $q=1/2$ at half filling the constant term $c_\al'$ tends to $c_{\al,\text{hom}}'$ as the number
  of spins tends to infinity. A similar analysis for the sextic chain (with $a=10^{-2}$) also
  shows a decrease in $|c_{\al,\text{hom}}-c_\al'|$ as $N$ increases in the same range, although
  the value of this difference is about three orders of magnitude higher than in the case of the
  Krawtchouk chain. This different behavior could be explained by the fact that the sextic chain
  is not invariant under $J_n\mapsto J_{N-n-2}$, as are the homogeneous and Krawtchouk chains. In
  fact, the latter two chains are the only XX spin chains of algebraic type (in the more general
  sense that the recursion coefficients $a_n$ are polynomial in $n$) with interactions $J_n$
  invariant under $n\mapsto N-n-2$ having a \emph{finite} total conformal length $\tilde\cL$.
\end{remark}
\begin{remark}
  The plots in Fig.~\ref{fig.SKr} suggest that the entanglement entropy of the Krawtchouk chain is
  invariant under $L\mapsto N-L$ not only for $q=1/2$ (at arbitrary filling), but also for
  arbitrary $q$ at half filling. That this is indeed the case can be deduced from a general
  property of the entanglement entropy, stemming from the fact that $\widehat J_n:=J_{N-n-2}=J_n$
  and $B_n$ is linear in $n$. Indeed, setting $B_n=B_0+b n$, with $B_0$ and $b$ independent of
  $n$, we have
  \[
    \widehat B_n:=B_{N-n-1}=2B_0+b(N-1)-B_n.
  \]
  From the previous equations for $\widehat J_n$ and $\widehat B_n$ we immediately obtain the
  following relation for the corresponding polynomials $\widehat P_n$:
  \[
    \widehat P_n(x)=(-1)^nP_n\Bigl(2B_0+b(N-1)-x\Bigr).
  \]
  This is readily seen to imply that
  \[
    \widehat\Phi_{nk}=(-1)^{n}\Phi_{n,N-k-1}\,,
  \]
  which yields the relation
  % \begin{align*}
  %   C_{mn}[\{0,\dots,&L-1\};\{0,\dots,M-1\}]\\
  %   &=(-1)^{m+n}C_{mn}[\{N-L,\dots,N-1\};\{N-M,\dots,N-1\}]\\
  %   &=(-1)^{m+n}C_{mn}[\{0,\dots,N-L-1\};\{N-M,\dots,N-1\}],
        %     \end{align*}
  \begin{multline*}
    \widehat C_{mn}[\{0,\dots,L-1\};\{0,\dots,M-1\}]=C_{mn}[\{N-L,\dots,N-1\};\{0,\dots,M-1\}]\\
                             =(-1)^{m+n}C_{mn}[\{0,\dots,L-1\};\{N-M,\dots,N-1\}],
                           \end{multline*}
  where the first argument denotes the block of spins considered and the second one the energy
  modes excited. Since the matrices $(A_{mn})$ and $((-1)^{m+n}A_{mn})$ are obviously similar, we
  deduce that
  \begin{align*}
      S_{\al}[\{0,\dots,N-L-1\};\{0,\dots,M-1\}]&=S_{\al}[\{N-L,\dots,N-1\};\{0,\dots,M-1\}]\\
                                              &=S_{\al}[\{0,\dots,L-1\};\{N-M,\dots,N-1\}]\,,
  \end{align*}
  where in the first equality we have applied the well-known invariance of the entanglement
  entropy under complements in position space. On the other hand, from the energy-position duality
  of the entanglement entropy~\cite{LYQ14,HA12,CFGT17} it follows that $S_\al$ is also invariant
  under complements in energy space. We thus obtain the relation
  \begin{equation}\label{SsymmM}
    S_{\al}[\{0,\dots,N-L-1\};\{0,\dots,M-1\}]= S_{\al}[\{0,\dots,L-1\};\{0,\dots,N-M-1\}].
  \end{equation}
  In particular, this implies that at half filling $S_\al$ is invariant under $L\mapsto N-L$, as
  claimed. Of course, for the Krawtchouk chain with $q=1/2$ we can combine Eqs.~\eqref{Ssymm}
  and~\eqref{SsymmM} to deduce that $S_\al$ is also invariant under $M\mapsto N-M$ (this also
  follows from a standard duality argument). Note, finally, that since the couplings of the sextic
  chain with $a\gg1$ are approximately symmetric under $n\mapsto N-n-2$, and its magnetic field
  term $B_n$ is linear in $n$, Eq.~\eqref{SsymmM} is approximately valid also in this case.
\end{remark}
\subsection{Ground state energy}
In Ref.~\cite{MSR21} it is conjectured that as $N\to\infty$ the ground state energy $E_0(N)$ of
an inhomogeneous XX chain~\eqref{Hchain} with $B_n=0$ for all $n$ behaves as
\begin{equation}\label{E0Nas}
  E_0(N)=-c_0\Si_N-c_B(J_0+J_{N-2})-\frac{\pi v_F}{24\wt\cN}+O(N^{-2})\,,
\end{equation}
where $\wt\cN=\tcL/\De x$,
\[
  \Si_N:=\sum_{n=0}^{N-2}J_n\,,
\]
and $c_0$, $c_B$, $v_F$ are three constants (representing the bulk energy per site, the boundary
energy and the Fermi velocity) which in the homogeneous case take the respective values $2/\pi$,
$4/\pi-1$, and $2$~\cite{Ca84,BCN86}. In the absence of a magnetic field the ground state is the
half-filled state~\eqref{halff}, whose energy $E_0(N)$ is given by Eq.~\eqref{E0N}. This quantity
can be \emph{exactly} computed for the Krawtchouk chain with $q=1/2$, since its single-particle
energies (after subtraction of the constant magnetic field $B_n=(N-1)/2$) are given by the formula
\[
  \vep_k=k-\frac12\,(N-1)\,,\qquad 0\le k\le N-1\,,
\]
whence
\begin{equation}\label{E0NKr}
  E_0(N)=-\frac{N^2}8\,.
\end{equation}
We shall next compare this exact value for $E_0(N)$ with its conjectured asymptotic
expansion~\eqref{E0Nas}, which by Eqs.~\eqref{JKraw} and \eqref{txtLKr} reads in this case
\begin{equation}\label{E0NasKr}
  E_0(N)=-c_0\Si_N-c_B\sqrt{N-1}-\frac{v_F}{24N}+O(N^{-2})\,.
\end{equation}
Using the exact value~\eqref{E0NKr} for $E_0(N)$ we deduce that in this case
\begin{equation}\label{asexpE0}
  c_0\Si_N=\frac{N^2}8-c_B\sqrt{N-1}-\frac{v_F}{24N}+O(N^{-2})\,.
\end{equation}
On the other hand, the leading asymptotic behavior as $N\to\infty$ of the sum
\[
  \Si_N=\frac12\sum_{n=0}^{N-2}\sqrt{(n+1)(N-n-1)}=\frac12\sum_{n=1}^{N-1}\sqrt{n(N-n)}
\]
can be determined from the Euler--Maclaurin formula~\cite{Kn51}
\begin{align*}
  2\Si_n-\int_1^{N-1}g(x)\,\diff x&=\frac12\,[g(N-1)+g(1)]+\frac12\,[g'(N-1)-g'(1)]+R_3\\
  &= g(1)-g'(1)+R_3
    =\sqrt{N-1}-\frac{N/2-1}{\sqrt{N-1}}+R_3\,,
\end{align*}
where $g(x):=\sqrt{x(N-x)}$. The remainder $R_3$ can be estimated as
\begin{align*}
  |R_3|&\le\frac{\ze(3)}{4\pi^3}\int_1^{N-1}|g'''(x)|\diff x
         =\frac{\ze(3)}{2\pi^3}\int_1^{N/2}g'''(x)\diff x
         =\frac{\ze(3)}{2\pi^3}\,[g''(N/2)-g''(1)]\\
       &=\frac{\ze(3)}{8\pi^3}\,N^2\big[(N-1)^{-3/2}-8N^{-3}\big]=O(N^{1/2})\,,
\end{align*}
where $\ze(s)$ denotes Riemann's zeta function. We also have
\[
  \int_0^Ng(x)\,\diff x=N^2\int_0^1\sqrt{t(1-t)}\,\diff t=\frac{\pi N^2}8\,,
\]
and hence
\[
  \int_1^{N-1}g(x)\,\diff x=\frac{\pi N^2}8-2\int_0^1\sqrt{x(N-x)}\,\diff x=\frac{\pi
    N^2}8+O(N^{1/2})\,.
\]
We thus conclude that
\[
  \Si_N=\frac{\pi N^2}{16}+O(N^{1/2})\,,
\]
in agreement with the right-hand side of Eq.~\eqref{asexpE0} if we take $c_0=2/\pi$ (as in the
homogeneous case).

It can be shown that the higher-order corrections in the Euler--Maclaurin formula are all
$O(N^{1/2})$, so that they cannot be used to compute $c_B$ and $v_F$ in closed form (this is
essentially due to the fact that the derivatives of $g(x)$ diverge at $x=0,N$). On the other hand,
the parameter $c_B$ can be computed through the formula
\[
  c_B=\lim_{N\to\infty}\frac1{\sqrt{N-1}}\,\bigg(\frac{N^2}8-\frac{2\Si_N}\pi\bigg)\,.
\]
By evaluating the right-hand side for large values of $N$ we have verified that this limit indeed
exists, and that $c_B=0.1323449$ to seven decimal places. Note that this value is about one half
of the corresponding one for the homogeneous case.

Using the previous estimate for the constant $c_B$, we have studied the behavior of the difference
$N^2/8-2\Si_N/\pi-c_B\sqrt{N-1}$
\begin{figure}[t]
  \centering
  \includegraphics[width=.55\textwidth]{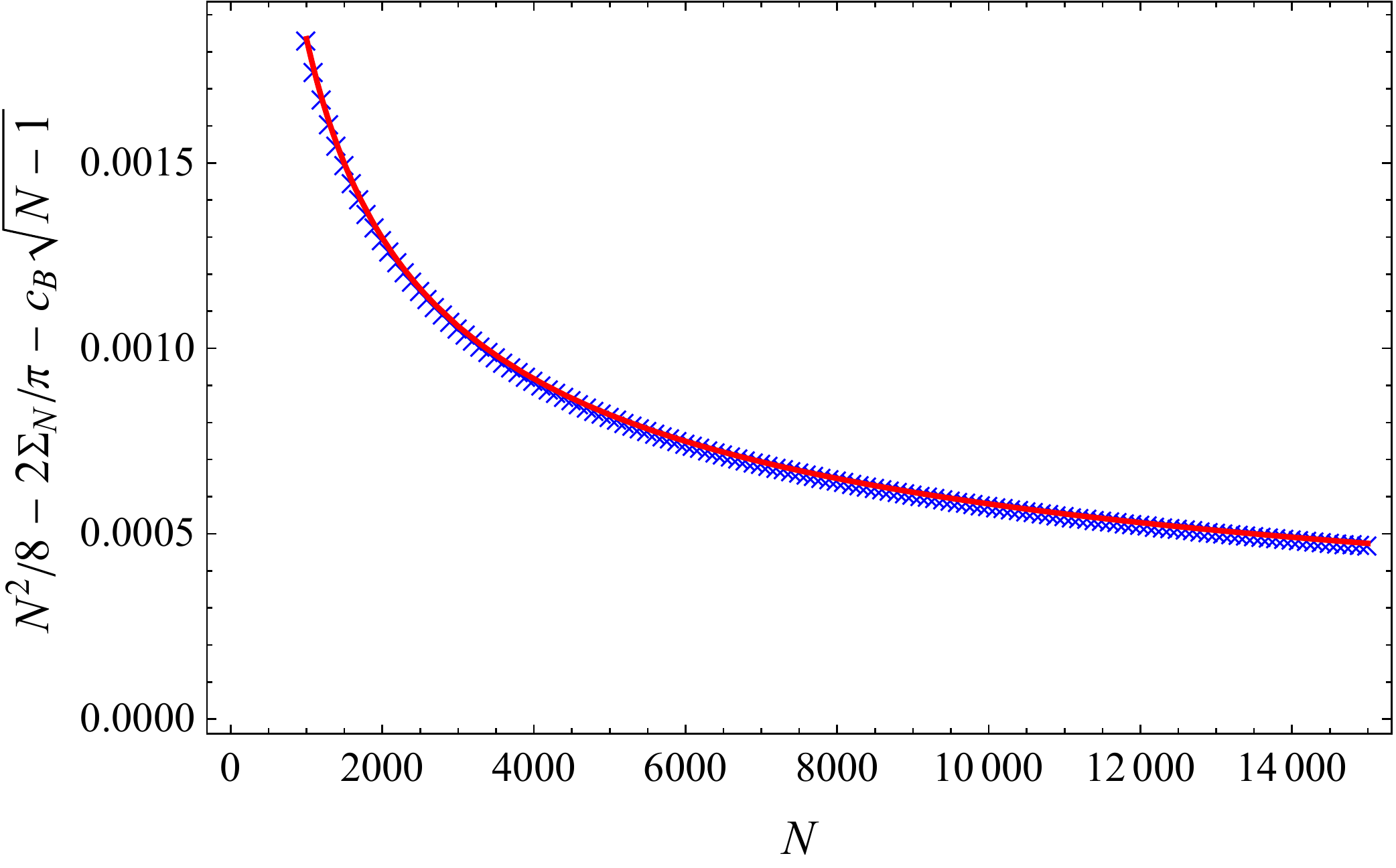}
  \caption{Difference $N^2/8-2\Si_N/\pi-c_B\sqrt{N-1}$ with $c_B=0.1323449$ for $N$ in the range
    $[1000,15000]$ in increments of 100 (blue crosses), compared to the curve $0.0580365/\sqrt{N}$
    (red line).}
  \label{fig.SiN}
\end{figure}
for $1000\le N\le 15000$, obtaining very strong numerical evidence that it is of order $N^{-1/2}$
instead of $N^{-1}$, as predicted by Eq.~\eqref{asexpE0} (cf.~Fig.~\ref{fig.SiN}). In other words,
our results suggest that for the Krawtchouk chain (with $q=1/2$) Eq.~\eqref{E0NasKr} should be
replaced by
\[
  E_0(N)=-c_0\Si_N-c_B\sqrt{N-1}-\frac{k}{\sqrt N}+o(N^{-1/2})\,,
\]
with $k\simeq0.0580365$.

\section{The Lamé chain}\label{sec.Lame}

We shall next study the chain associated to the quantum (finite gap) Lamé potential introduced in
Ref.~\cite{FG20}, whose parameters are given by
\begin{equation}
  \label{LameJB}
  J_n=\sqrt{(n+1)(N-n-1)(n+1/2)(N-n-3/2)}\,,\qquad B_n=0\,.
\end{equation}
Although an approximation to the entanglement entropy of this chain at half filling was obtained
in Ref.~\cite{FG20}, for the sake of consistency we shall next outline the derivation of an
equivalent formula using the present approach and notation. To this end, we first write
\[
  J_n=(N-1)^2\sqrt{\bigg(\frac{x_n}\cL+\vep\bigg)\bigg(1-\frac{x_n}\cL\bigg)
    \bigg(\frac{x_n}\cL+\frac\vep2\bigg)\bigg(1-\frac{x_n}\cL-\frac{\vep}2\bigg)}\,,
  \qquad\vep:=\frac1{N-1}\,,
\]
and thus (up to an irrelevant constant factor)
\begin{equation}
  \label{JxLame}
  J(x)=\sqrt{p(x/\cL)}\,,\qquad p(\xi):=(\xi+\vep)(\xi+\vep/2)(1-\xi-\vep/2)(1-\xi)\,.
\end{equation}
Note that in this case we cannot just take $\vep=0$, since the integral of $1/J(x)$ would then
diverge at $x=0,\cL$. It is also worth mentioning that in this case we cannot extend the range of
$\xi$ all the way to $1$, since $p(\xi)$ is negative for $\xi>1-\vep/2$. To evaluate the
integral~\eqref{mainint} we perform the change of variable $\xi=(z+1-\vep)/2$, obtaining
\begin{align*}
  \int_0^s\frac{\diff\xi}{\sqrt{(\xi+\vep)(\xi+\frac\vep2)(1-\xi-\frac\vep2)(1-\xi)}} &=
                                                                                        2k\int_{-1+\vep}^{2s-1+\vep}\frac{\diff z}{\sqrt{(1-z^2)(1-k^2z^2)}}\\
                                                                                      &=2k\Big[F\Bigl(\arcsin(2s-1+\vep)\Bigr)+F\Bigl(\arcsin(1-\vep)\Bigr)\Big],
\end{align*}
where the modulus of the elliptic integral is
\[
  k:=(1+\vep)^{-1}=1-\frac1N<1\,.
\]
We thus have
\[
  \tx(x)=2k\cL\Big[F\Bigl(\arcsin(2\xi-1+\vep)\Bigr)+F\Bigl(\arcsin(1-\vep)\Bigr)\Big],\qquad
  \xi:=x/\cL\,.
\]
In order to compute the chain's conformal length~$\tcL$ we must specify the upper limit of the
variable $\xi$, which in this case cannot be extended to $1$ for the reason explained above. In
fact, although the integrand in Eq.~\eqref{mainint} remains real up to $\xi=1-\vep/2$, it is more
convenient (and of no consequence in the limit $N\to\infty$) to use the symmetric interval
$-1+\vep\le\xi\le1-\vep$. With this choice we obtain
\[
  \tcL=4k\cL F\Bigl(\arcsin(1-\vep)\Bigr),
\]
from which it follows that
\[
  \sin\left(\frac{\pi\tell}{\tcL}\right)=\cos\left(\frac{\pi F\left(\arcsin(2\la-1+\vep)\right)}%
    {2F\left(\arcsin(1-\vep)\right)}\right).
\]
For $0<\la<1$ the numerator of the argument of the cosine tends to the finite limit
$F(\arcsin(2\la-1),1)=\arctanh(2\la-1)$ as $N\to\infty$ (i.e., $\vep\to0+$), while the denominator
tends to $K(1)=+\infty$. Thus for sufficiently large $N$ we can take the function $f$ in
Eq.~\eqref{fnonh} simply as
\begin{equation}\label{fLame}
  f(N,\la)\simeq\frac{N\tcL/\cL}{\pi}\,J(\ell)\simeq\frac{4N}\pi\,F\Bigl(\arcsin(1-\vep)\Bigr)
  \la(1-\la),
\end{equation}
up to lower-order terms in $N$. Note that $f(N,\la)$ is invariant under $\la\mapsto 1-\la$, which
is consistent with the symmetry of the coupling~\eqref{LameJB} with respect to the chain's
midpoint. It can be shown that in the $N\to\infty$ limit we have
\[
  F\Bigl(\arcsin(1-\vep)\Bigr)=\frac12\log(2N)+\kappa+o(1)\,,
\]
with $\kappa=0.1882264\dots$, so that $f(N,\la)\sim N\log N$ in this limit. Thus when $N\to\infty$
with $\la$ fixed the entanglement entropy of the Lamé chain~\eqref{LameJB} at half filling should
behave as
\begin{equation}
  \label{SalLame}
  S_\al(N,\la)=\frac12\,(1+\al^{-1})\log[N\la(1-\la)]+O\bigl(\log(\log N)\bigr).
\end{equation}
Equation.~\eqref{fLame} is in agreement up to lower order terms with the analogous equation in
Ref.~\cite{FG20}, obtained by replacing $F(\arcsin(1-\vep))$ by $K$, since~\cite{OLBC10}
\[
  K\Bigl((1+\vep)^{-1}\Bigr)=\frac12\log(8N)+o(1).
\]
\begin{figure}[t]
  \includegraphics[height=.32\textwidth]{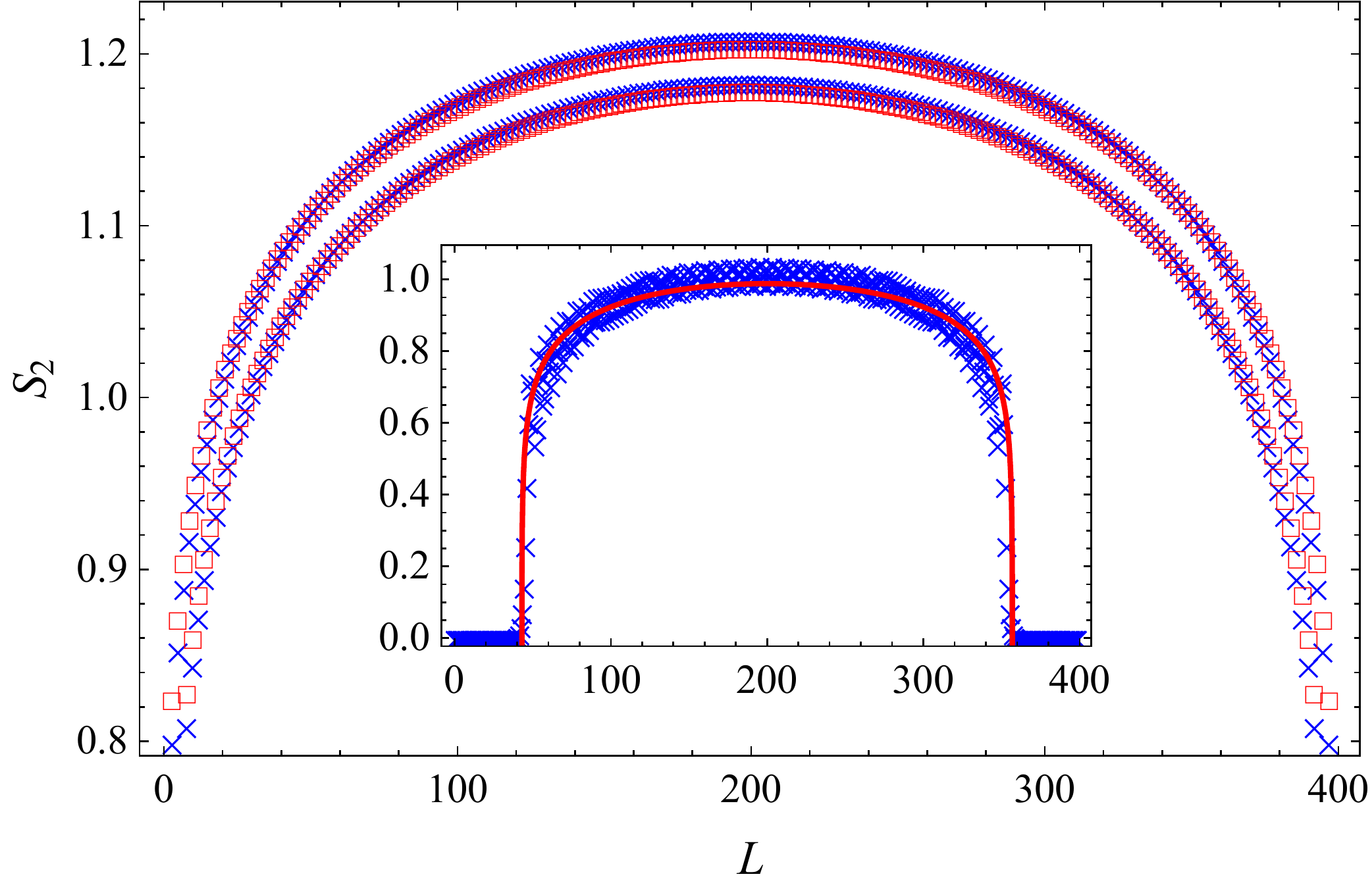}\hfill
  \includegraphics[height=.32\textwidth]{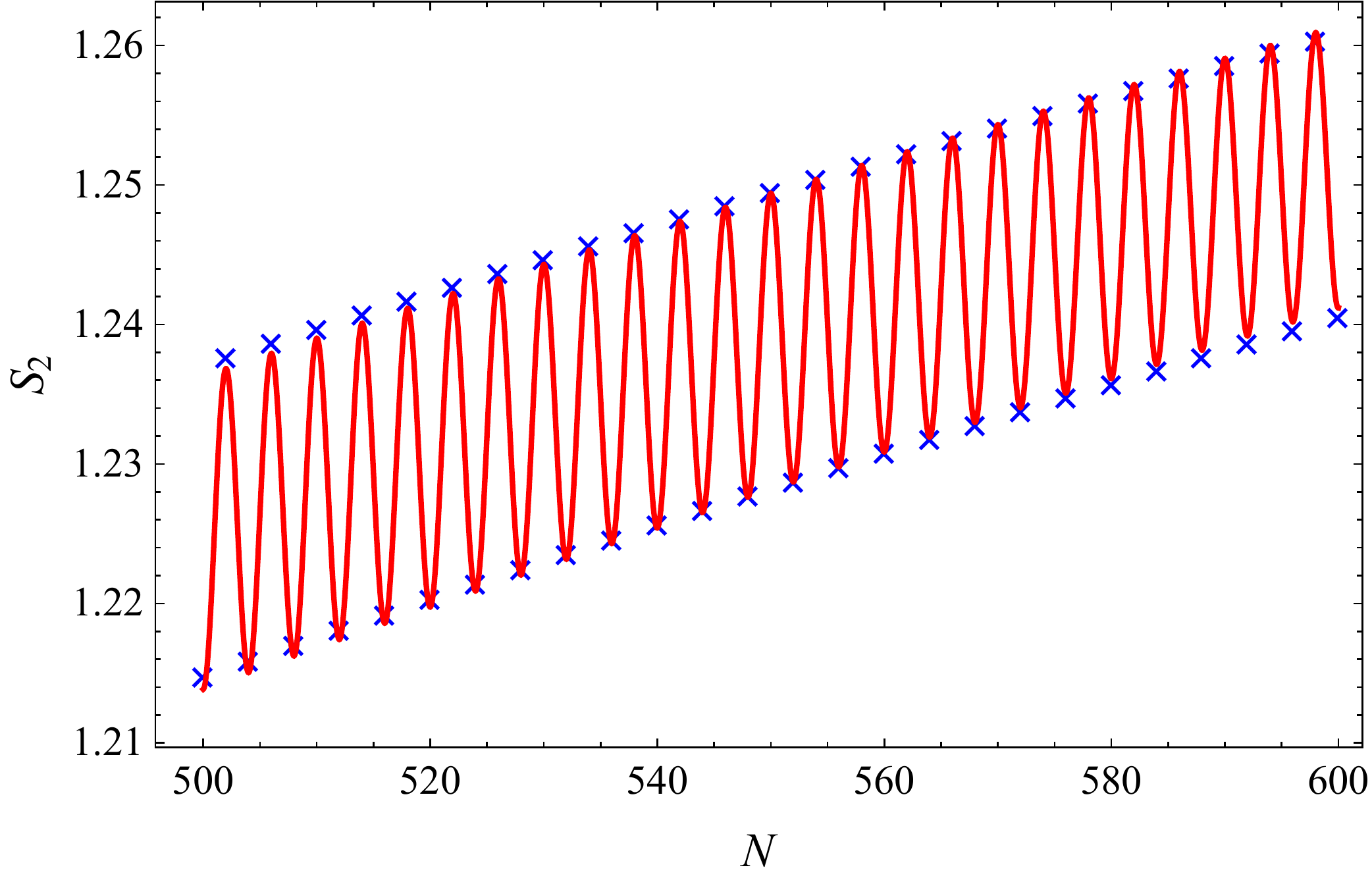}
  \caption{Left: Rényi entanglement entropy with parameter $\al=2$ for the Lamé chain with $N=400$
    spins at half filling (blue crosses) compared to its asymptotic
    approximation~\eqref{Sasymposc}-\eqref{fLame} (red squares). The inset shows the analogous
    plot for the Fermi momentum $k_F=\pi/4$ compared to the heuristic
    approximation~\eqref{Salbe}-\eqref{fkFL} (red line). Right: Rényi entanglement entropy $S_2$
    for the Lamé chain with $\la=1/2$ for an even number of spins $N$ in the interval $[500,600]$
    at half filling (blue crosses) compared to its approximation~\eqref{Sasymposc}-\eqref{fLame}
    with $(-1)^L$ replaced by $\sin((2L+1)\pi/2)$ (red line).}
  \label{fig.Lame}
\end{figure}%
It should also be noted that the logarithmic divergence of $f(N,\la)/N$ as $N\to\infty$ with $\la$
fixed is due to the presence of double zeros of $J(x)$ at both endpoints of the interval $[0,\cL]$
in this limit, and can thus only happen when the polynomial $p(\xi)$ in Eq.~\eqref{pdef} is of
degree four.

We have numerically checked that the asymptotic formula~\eqref{Sasymposc}-\eqref{fLame} still
provides a reasonable approximation to the entanglement entropy at half filling in this case,
though not as precise as for the sextic and Krawtchouk chains. In particular, the oscillating term
proportional to $f(N,\la)^{-1/\al}$ in Eq.~\eqref{Sasymposc} reproduces with acceptable accuracy
the parity oscillations in $S_\al$ present when $\al\ge1$ (see, e.g., Fig.~\ref{fig.Lame} (left)
for $\al=2$ and $N=400$ spins). As before, at Fermi momenta $k_F\ne\pi/2$ the Rényi entanglement
entropy $S_\al$ is virtually zero over two intervals of the form $[1,L_1]$ and $[N-L_1,N-1]$, and
the asymptotic formula~\eqref{Sasymposc}-\eqref{fLame} fails. However, proceeding as above we can
derive a rough heuristic approximation to $S_\al$ in the interval $[L_1+1,N-L_1-1]$ using
Eq.~\eqref{Salbe} with
\begin{equation}\label{fkFL}
  f(N,\la)=\frac{4(N-2L_1)}\pi\,F\Bigl(\arcsin(1-\vepeff),(1+\vepeff)^{-1}\Bigr)
  \laeff(1-\laeff)\sin k_F\,,
\end{equation}
with $\laeff$ given by Eq.~\eqref{laeff} with $L_2=N-L_1$ and $\vepeff=(N-2L_1-1)^{-1}$.

Since the dependence on the number spins $N$ of the asymptotic
formula~\eqref{Sasymposc}-\eqref{fLame} for the entanglement entropy at half filling is
nontrivial, it is also of interest to study the growth of $S_\al(N,\la)$ with $N$ for fixed values
of $\la$. We have checked that the latter formula captures the behavior of $S_\al(N,\la)$ with
great accuracy, and in particular reproduces the parity oscillations that appear when $\al\ge1$
(see, e.g., Fig.~\ref{fig.Lame} (right) for the case $\la=1/2$ and even $N$ in the
interval~$[500,600]$). Remarkably, although the two parameters $c_\al'$ and $\mu_\al$ appearing in
Eq.~\eqref{Sasymposc} are fitted, they turn out to be of the same order of magnitude as their
counterparts~\eqref{cpal}-\eqref{mutheor} for the homogeneous XX chain.
%
% Changed
%
Note, however, that in this case it should not be expected that the constant term $c_{\al}'$ tend
to $c_{\al,\text{hom}}'$ as $N\to\infty$, since the subleading term in the asymptotic expansion of
$S_\al$ is no longer constant but of the order of $\log(\log N)$ (cf.~Eq.~\eqref{SalLame}).

\section{Conclusions and outlook}\label{sec.conc}

In this work we study a large class of inhomogeneous XX spin chains whose squared couplings are a
polynomial of degree at most four in the site index. This class includes some previously studied
models related to classical Krawtchouk and dual Hahn polynomials~\cite{CNV19}, as well as the
inhomogeneous XX chains related to QES models on the line classified in Ref.~\cite{FG20}. We show
how to exactly compute the leading term in the asymptotic expansion of the block entanglement
entropy of these models (in a constant magnetic field at half filling) from their continuum limit,
which coincides with the CFT of a massless Dirac fermion in a curved ($1+1$)-dimensional
background~\cite{DSVC17,RDRCS17,TRS18}. We next focus on three inhomogeneous chains with algebraic
interactions, associated to the sextic oscillator QES potential, the Krawtchouk polynomials and
the periodic quantum Lamé potential. We exploit the relation of XX chains with finite families of
orthogonal polynomials to numerically compute the Rényi entanglement entropy of the latter chains
for a large number of spins. When the Rényi parameter~$\al$ is less than one we find that, as
expected, the asymptotic formula reproduces with great accuracy the behavior of the entanglement
entropy. On the other hand, for $\al\ge1$ the Rényi entropy presents parity oscillations whose
amplitude increases with $\al$, as is known to be the case for the homogeneous XX
chain~\cite{FC11}. We show that these oscillations are reproduced with excellent accuracy by the
Fagotti--Calabrese formula for the homogeneous chain~\cite{FC11}, replacing the block's and the
chain's length by their conformal counterparts.
%
% Changed
%
In fact, for the sextic and Krawtchouk chains (at half filling and in a constant magnetic field)
we have found rather compelling numerical evidence that the subleading non-universal (constant)
term in the asymptotic expansion of the entanglement entropy tends to its counterpart for the
homogeneous XX chain as the number of spins tends to infinity. We conjecture that this is actually
the case for the class of algebraic chains studied in this paper when the chain's conformal length
is finite, i.e., when the squared couplings considered as functions of the site index $n$ have no
multiple real roots in the interval $[0,N]$.

All of the above results apply to the case of half filling and constant magnetic field, which is
the one usually considered in the literature in the inhomogeneous case. In this work we have also
studied in some detail the non-standard situation of arbitrary filling and/or inhomogeneous
magnetic field. We have found that the main difference with the standard situation is that the
block entanglement entropy vanishes when the block's length is either small or close to the
chain's length. Thus at fillings other than one-half, or in an inhomogeneous magnetic field, the
first few and last spins in the chain become disentangled from the rest. This is in fact one of
the paper's main results, which certainly deserves further theoretical analysis. Another
interesting feature of the non-standard case is that the oscillations of the entropy when
$\al\ge1$ are considerably more complex than in the standard one, and in particular are not well
described by a modification of the Fagotti--Calabrese formula along the lines mentioned above.

One of the models studied in this paper, namely the Krawtchouk chain
(cf.~Section~\ref{sec.Krawtchouk}), has the rather unusual property that its single-particle
energies at zero magnetic field can be exactly computed (they are simply the numbers $-(N-1)/2+k$,
with $0\le k\le N-1$). This of course makes it trivial to evaluate the ground-state energy in
closed form for an arbitrary number of spins $N$. We have compared this exact result with the
recently proposed asymptotic expansion in Ref.~\cite{MSR21}, finding that they match only to
leading order.

The above results clearly suggest several avenues for future research that we shall now briefly
outline.
%
% Changed
%
To begin with, we would like to find a theoretical justification of the fact that the constant
term in the asymptotic expansion of the entanglement entropy of chains with algebraic interactions
and finite conformal length (at half filling and in a constant magnetic field) seems to coincide
with the analogous term for the homogeneous XX model in the limit of large $N$. An outstanding
open problem of considerable interest is that of deriving an asymptotic formula for the leading
behavior of the entanglement entropy in the non-standard scenario of arbitrary filling and/or
inhomogeneous magnetic field. Our numerical results show that any such formula must necessarily
vanish when the block length is either small or close to the chain's length. Likewise, it would
also be of interest to find a formula describing the entropy's complex oscillations when the Rényi
parameter is greater than or equal to one, akin to the Fagotti--Calabrese formula for the
homogeneous chain. As mentioned in the Introduction, in the homogeneous case the behavior of the
multiblock entanglement entropy has been thoroughly analyzed (see, e.g.,
\cite{CH09,CCT09,ATC10,AEF14,CFGT17}). Again, the generalization of some of these results to the
inhomogeneous case would certainly be worth exploring. Finally, another problem suggested by the
present work is to understand why the asymptotic formula in Ref.~\cite{MSR21} for the ground-state
energy of inhomogeneous XX spin chains at zero magnetic field fails to reproduce the subleading
behavior of the Krawtchouk chain, and how it should be modified to account for this model and
similar ones.

\appendix

\section{Reduction of the integral~\eqref{mainint} to Legendre canonical form}

In this appendix we present a simplified procedure, based on the classical one described in
Ref.~\cite{La89}, for reducing the integral~\eqref{mainint} to canonical form in the nontrivial
case in which all the roots of the third- or fourth-degree polynomial $p(\xi)$ in Eq.~\eqref{pdef}
are simple.

To begin with, we can assume w.l.o.g.\ that $p(\xi)$ is of degree four, since if $\deg p=3$ the
projective change of variable $\xi=\xi_0+1/z$, where $p(\xi_0)\ne0$, transforms $p$ into a fourth
degree polynomial. We can thus write
\begin{equation}\label{pfac1}
  p(\xi)=\nu_0p_1(\xi)p_2(\xi)\,,
\end{equation}
with $\nu_0\ne0$ and
\begin{equation}\label{pfac2}
  p_i(\xi):=\xi^2+2\al_i \xi+\be_i\,,\qquad i=1,2\,.
\end{equation}
We next show that it is always possible to find a projective change of variable~\eqref{proj} with
$c=1$ transforming the product $p_1(\xi)p_2(\xi)$ into the polynomial
\begin{equation}\label{canformp}
  \hat p(z)=(A_1z^2+B_1)(A_2z^2+B_2)\,,
\end{equation}
where $A_i,B_i\in\RR$. Indeed, such a change of variable maps each $p_i(\xi)$ into the polynomial
\[
  \hat p_i(z)=A_iz^2+2C_iz+B_i
\]
with
\[
  A_i=\De^{-1}p_i(a)\,,\qquad C_i=\De^{-1}\big[ab+\al_i(ad+b)+\be_id\big]\,,\qquad
  B_i=\De^{-1}\big[b^2+2\al_ibd+\be_id^2\big]\,.
\]
Requiring that $C_1=C_2=0$ leads to the linear homogeneous system
\[
  \left(
  \begin{array}{cc}
    a+\al_1& \al_1a+\be_1\\
    a+\al_2& \al_2a+\be_2
  \end{array}
\right)
  \left(\begin{array}{cc}
    b\\d
  \end{array}\right)
  =0\,.
\]
The necessary and sufficient condition for the latter system to have a nontrivial solution is that
the determinant of its coefficient matrix vanish, i.e., that
\[
  (\al_2-\al_1)a^2+(\be_2-\be_1)a+\al_1\be_2-\al_2\be_1=0\,.
\]
For this quadratic equation in $a$ to have real roots its discriminant
\[
  \de:=(\be_1-\be_2)^2-4(\al_2-\al_1)(\al_1\be_2-\al_2\be_1)
\]
must be nonnegative. Calling $r_i^{1,2}$ the two (possibly complex) roots of $p_i(\xi)$, we can
rewrite $\de$ as
\[
  \de=\prod_{i,j=1}^2(r_1^i-r_2^j)\,.
\]
If $p$ has at least a pair of complex conjugate roots $u\pm\iu v$, say $r_2^1$ and $r_2^2$, the
previous expression reduces to
\[
  \de=\prod_{i=1}^2\big[(r_1^i-u)^2+v^2\big]\,.
\]
This is clearly positive if the roots $r_1^{1,2}$ are real, whereas when $r_1^{1,2}=s\pm\iu t$ we
have
\[
  \de=\left|(s-u+\iu t)^2+v^2\right|^2>0.
\]
On the other hand, if $p(\xi)$ has four distinct real roots we can assume w.l.o.g. (by redefining
the two factors $p_{1,2}$ if necessary) that $r_1^{1,2}$ are the two largest roots of $p$. Hence
also in this case $\de>0$, which concludes the proof of our claim. Note, finally, that since by
hypothesis the polynomial $p(\xi)$ has no multiple roots none of the coefficients in
Eq.~\eqref{canformp} can vanish, since otherwise the transformed polynomial $\hat p$ would have a
double root either at $0$ or at $\infty$.

\goodbreak\medskip\noi
I.\en $p$ has four simple real roots

\smallskip\noi If all the roots of $p(\xi)$ ---and, hence, of $\hat p(z)$--- are real then
$A_iB_i<0$ for $i=1,2$ in Eq.~\eqref{canformp}. Applying, if necessary, a dilation we can
therefore assume without loss of generality that
\[
  \hat p(z)=\nu_0(1-z^2)(1-Az^2)
\]
with $\nu_0\ne0$ and $A>0$. Since the projective transformation $z=1/w$ maps $\hat p(z)$ to the
polynomial
\[
  A\nu_0(1-w^2)(1-w^2/A)\,,
\]
we can also take $A=k^2$ with $0<k<1$. Thus in this case $p(\xi)$ can be reduced to the canonical
forms
\[
  \hat p_\pm(z)=\pm\nu(1-z^2)(1-k^2z^2)\,,\qquad 0<k<1\,,\quad\nu>0\,.
\]
The positivity intervals of $\hat p_+$ and $\hat p_-$ are respectively
$(-\infty,-1/k)\cup(-1,1)\cup(1/k,\infty)$ and $(-1/k,-1)\cup(1,1/k)$, although by the even
character of $\hat p_\pm$ we can restrict ourselves to the intervals $(-1,1)\cup(1/k,\infty)$ and
$(1,1/k)$.

For $\hat p_+$ and $z\in(-1,1)$ we apply the standard change of variable $z=\sin\th$ to obtain
\[
  \int_0^x\frac{\diff z}{\sqrt{(1-z^2)(1-k^2z^2)}}=F(\arcsin x,k)\,.
\]
The interval $(1/k,\infty)$ can be mapped to the standard one $(0,1)$ by the projective change of
variable $z=1/(kw)$. In this way ---or, equivalently, performing the change of variable
$z=(1/k)\csc\th\,$--- we obtain
\[
  \int_{1/k}^x\frac{\diff z}{\sqrt{(1-z^2)(1-k^2z^2)}}=K(k)-F(\arcsin(1/(kx),k)\,.
\]

Consider next the canonical form $\hat p_-(z)$ in the positivity interval $(1,1/k)$. Although the
change of variable
\[
  w=\frac1{\sqrt k}\,\frac{1-\sqrt{\tilde k}z}{1+\sqrt{\tilde k}z}\,,\qquad
  \text{with}\quad\tilde k:=\bigg(\frac{1-\sqrt k}{1+\sqrt k}\bigg)^2\,,
\]
maps $\hat p_+(z)$ into a positive multiple of $\hat p_-(w)$ with $k$ replaced by $\tilde k$, and
the interval $(-1,1)$ to $(1,1/\tilde k)$, it is easier in this case to perform the change of
variable
\[
  z=\frac1{k}\sqrt{1-{k'}^2\sin^2\th}\,,\qquad \text{with}\quad k'=\sqrt{1-k^2}\,,
\]
in the integral for $\hat p_-(z)$. In this way we obtain
\[
  \int^x_1\frac{\diff
    z}{\sqrt{(z^2-1)(1-k^2z^2)}}=K(k')
  -F\left(\arcsin\biggl(\frac{\sqrt{1-k^2x^2}}{k'}\,\biggr),k'\right).
\]

\goodbreak\medskip\noi
II.\en $p$ has two simple real and two complex conjugate roots

In this case we can take $A_1B_1<0$ and $A_2B_2>0$ in Eq.~\eqref{canformp}. We can thus write
(applying, if necessary, a dilation)
\[
  \hat p(z)=\nu_0(1-z^2)(A_2z^2+B_2)
\]
with $\nu_0\in\RR$. We can also assume w.l.o.g.\ that $A_2,B_2>0$, and set
\[
  A_2=k\sqrt{A_2^2+B_2^2}\,,\qquad B_2=k'\sqrt{A_2^2+B_2^2}
\]
with $0<k<1$ and $k'=\sqrt{1-k^2}$. Hence we can write
\[
  \hat p(z)=\pm\nu(1-z^2)(k'^2+k^2z^2)=:\hat p_\pm(z)\,,\qquad 0<k<1\,,\quad\nu>0\,.
\]
Moreover, the projective transformation $z=1/w$ maps $\hat p_-(z)$ into $\hat p_+(w)$, with $k$
and $k'$ interchanged. Thus in this case $p(\xi)$ can be reduced to the single canonical form
$\hat p_+(z)$, whose positivity interval is $(-1,1)$. The change of variable $z=\cos\th$ then
leads to the formula
\[
  \int_0^x\frac{\diff z}{\sqrt{(1-z^2)(k'^2+k^2z^2)}}=K(k)-F(\arccos x,k)\,.
\]

\goodbreak\medskip\noi
III.\en $p$ has four simple complex roots.

In this case we have $A_iB_i>0$ in Eq.~\eqref{canformp}. Moreover, since $\hat p$ must be positive
on some open interval we can take w.l.o.g.\ $A_i,B_i>0$. We can therefore write (up to a dilation)
\[
  \hat p(z)=\nu(1+z^2)(1+A^2z^2)
\]
with $\nu,A>0$. Applying, if needed, a projective transformation $z=1/w$, we can assume w.l.o.g.\
that $A<1$, and thus set $A={k'}^2=1-k^2$ with $0<k<1$. Hence in this case $p(\xi)$ can be reduced
to the canonical form
\[
  \hat p(z)=\nu(1+z^2)(1+{k'}^2z^2)\,,\qquad 0<k<1\,,\quad\nu>0\,,
\]
which is positive everywhere. Performing the change of variable $z=\tan\th$ we then obtain
\[
  \int_0^x\frac{\diff z}{\sqrt{(1+z^2)(1+{k'}^2z^2)}}=F(\arctan x,k)\,.
\]

\acknowledgments This work was partially supported by grants PGC2018-094898-B-I00 from Spain's
Mi\-nis\-te\-rio de Ciencia, Innovaci\'on y Universidades and~G/6400100/3000 from Universidad
Complutense de Madrid. The authors would like to thank Bego\~na Mula, Silvia N. Santalla and
Javier Rodr\'\i guez Laguna for helpful discussions, and an anonymous referee for his suggestions.

% \bibliographystyle{JHEP}
% \bibliography{cmprefs}

\providecommand{\href}[2]{#2}\begingroup\raggedright\endgroup

\end{document}